\documentclass[12pt]{iopart}

\usepackage{graphicx}
\usepackage{subfigure}
\usepackage{algorithm}
\usepackage{algorithmic}
\usepackage{cite}
\usepackage{url}
\usepackage{iopams}   
\usepackage{CJK}
\usepackage{multirow}
\usepackage{bm}
\usepackage{ulem}
\usepackage{pdfpages}

\begin{document}

\title[]{Identifying polycentric urban structure using the minimum cycle basis of road network as building blocks}

\author{Yuanbiao Li$^{1,2}$, Tingyu Wang$^{1,2}$, Yu Zhao$^{1,2}$, Bo Yang$^{1,2,*}$}
\address{$^1$ Data Science Research Center, Kunming University of Science and Technology, Yunnan, Kunming 650500, People's Republic of China}
\address{$^2$ Faculty of Science, Kunming University of Science and Technology, Yunnan, Kunming 650500, People's Republic of China}

\address{$^*$ Author to whom any correspondence should be addressed.}
\ead{yangbo@kust.edu.cn}

\vspace{10pt}
\begin{indented}
\item[]April 2024
\end{indented}

\begin{abstract}
In a graph, the minimum cycle bases are a set of linearly independent cycles that can be used to represent any cycle within that cycle space of graph. These bases are useful in various contexts, including the intricate analysis of electrical networks, structural engineering endeavors, chemical processes and surface reconstruction techniques etc. This study focuses on six cities in China to explore the topological characteristics, the centrality of nodes and robustness of urban road networks based on motif and minimum cycle bases. Some interesting conclusions are obtained: the frequency of motifs containing cycles exceeds that of random networks with equivalent degree sequences; the frequency distribution of minimum cycle's length and surface areas obey the power-law distribution. The cycle contribution rate is introduced to investigate the centrality of nodes within road networks, and has a significant impact on the total number of cycles in the robustness analysis. Finally, we construct two types of cycle-based dual networks for urban road networks by representing cycles as nodes and establishing edges between two cycles sharing a common node and edge respectively. The results show that cycle-based dual networks exhibit small-world and scale-free properties.
\end{abstract}
\vspace{2pc}
\noindent{\it Keywords}: urban road network, motif, the minimum cycle, power-law distribution
%
%

\section{Introduction}\label{Indro}
In recent years, researchers have embarked on a comprehensive exploration of various transportation networks, including railways\cite{r1,r2}, aviation\cite{r3,r4}, urban roads\cite{r5,r6, r7, r8,r9,r10,r11,r12,r13,r14,r15,r16,r17}, and others\cite{r18,r19,r20}. As a vital component of a city, urban roads play an important role in the normal operation of urban functions\cite{r21,r22}. However, with the acceleration of urbanization and the rapid surge in transportation demand, the complexity of urban road systems has increased significantly\cite{r23,r24}, and complex network theory has gained widespread acceptance in observing and analyzing this intricate structure\cite{r25,r26,r27}.
In the study of urban road networks, researchers have explored the structural characteristics of urban roads from various perspectives, including self-similarity\cite{r5}, connectivity\cite{r6}, spatial layout\cite{r7}, growth and evolution\cite{r8}, topological structure\cite{r9,r10,r11,r12,r13}, and robustness\cite{r14,r15,r16,r17}.

Regarding the topological structure of urban road networks, Zhang et al.\cite{r9,r10}studied the relationship between morphological features and topological characteristics such as network degree centrality, network betweenness centrality, and network closeness centrality. Their findings revealed that betweenness centrality can effectively distinguish and describe different morphological road networks. Tsiotas et al.\cite{r11} analyzed the relationship between urban roads and space, discovering that the degree distribution and connectivity of road networks are influenced by urban spatial constraints. Jiang et al.\cite{r12} studied the degree characteristics of urban road networks and found a lack of degree correlation within the road networks. Shang et al.\cite{r13} examined the topological structure of urban road networks, including clustering coefficient and average path length, and found that the small-world characteristics of urban road networks are not significant.

Regarding the robustness of urban road networks, several key observations can be made. Firstly, the adoption of different attack strategies has a significant impact on network robustness. Liu et al.\cite{r14} studied the model of cascading failures in urban road networks, revealing that road intersections have the greatest impact on robustness. Furthermore, random attacks have a weaker impact on network robustness compared to targeted attacks. Secondly, there are differences in network robustness across different granularities. Duan et al.\cite{r15} modeled urban road networks based on three granularities: segment-based, stroke-based, and community-based. Their results showed that due to the similarity in degree distribution and topological structure among urban road networks, robustness tends to be consistent. However, structural differences across granularities lead to variations in robustness. Thirdly, the impact of attacks targeting different topological metrics on network robustness varies. Zhao et al.\cite{r16} employed a duality method\cite{r17}(which abstracts road segments  as nodes in the road network and converts connectivity relationships, such as intersections, into edges in the network) to transform urban road networks into topological graphs for robustness analysis. Their findings indicated that node degree has the greatest impact on robustness, followed by node betweenness and edge betweenness.

Milo et al.\cite{r28,r29} proposed the concept of network motifs, which are small subgraphs of a network formed by a limited number of nodes arranged in a specific topological structure. In the context of road networks, motifs refer to road structural units with specific spatial configurations and traffic characteristics.
These motifs appear more frequently in urban road networks than in random networks with the same degree sequence. The design and layout of these motifs directly impact the traffic flow efficiency of the road network. An optimal layout of motifs can make the road network more compact and connected, enhancing the accessibility and convenience of the city\cite{r30}.

Currently, limited research has been conducted on urban road networks utilizing motif methods, and the Z-scores of different motifs remain unexplored\cite{r30,r31,r33}. To address this gap, we conducted a detailed analysis of urban road networks using motif methods and obtained the Z-scores for various motifs. The results revealed that motifs with cycles exhibited positive Z-scores, indicating that their frequency of occurrence was significantly higher than in random networks with the same degree sequence. Based on this finding, this study proposes an analysis of urban road networks, treating the minimum cycle basis as the fundamental units.

As an essential component of networks, cycles have received widespread attention including algorithm and application of minimum cycle base\cite{r34,r35,r36,r37}, the distribution of cycles of different sizes in real networks and artificial networks\cite{r38,r39,r40,r41,r42,r43,r44}, etc. Zhang et al.\cite{r45} proposed Cycle Nodes Ratio (CNR) for network classification. Fan et al.\cite{r46} proposed the use of the cycle ratio as a metric to assess the importance of nodes, comparing it with the results obtained from degree, H-index, and coreness. Experiments on real-world networks suggest that the cycle ratio contains rich information in addition to well-known benchmark indices. But we measured the importance of nodes based on their cycle ratio in road network. The results revealed that nodes with higher cycle degrees within larger cycles tended to have higher cycle ratio values, which did not effectively identify the central nodes in the road network. To address this issue, we introduced the concept of cycle contribution rate to assess node importance, considering both the number and length of cycles associated with each node. To compare the impact of removing node random and according to degree, betweenness, cycle degree, cycle ratio and cycle contribution rate on network robustness, we proposed the average length of cycles and the number of cycles based on the relative size of the largest connected subgraph to measure the variation of cycles in the road network.

Finally, we constructed two types of cycle-based dual networks by treating the minimum cycle basis as nodes and establishing connections based on shared one node and one edge between cycles respectively. Then we analyzed the topological properties of the cycle-based dual network, including degree distribution, clustering coefficient, average path length and network diameter.

The structure of this paper is arranged as follows: In Section \ref{S2}, we provide a detailed introduction to the minimum cycle basis model, robustness analysis methods, and the cycle-based dual network model. In Section \ref{S3}, we present a comprehensive analysis and further discussion of the research findings in urban road networks. Finally, in Section \ref{S4}, we summarize and draw some conclusions.

\section{Model and Method}\label{S2}

\subsection{Methods to Calculate the Cycle Features}
For a given undirected network $G=(V, E)$, where $V=\{{v_{1}, v_{2},~...~,v_{N}}\}$ represents the set of $N$ nodes, and $E$ represents the set of edges which are unordered pairs of elements of $V$. A cycle $O_{i}: v_{1}v_{2} ... v_{k-1}v_{k}v_{1}$ is a subgraph such that every vertex has even degree, and its length is the number of its edges. Cycles in graph generate the cycle space of $G$. A cycle basis of $G$ is defined as a maximal set of linearly independent cycles. The dimension of the cycle space is $M-N+\kappa(G)$, where $\kappa(G)$ is the number of connected components of $G$. For an unweighed graph, the cycle basis where the sum of the length of the cycles is minimum are called a minimum cycle basis of $G$ \cite{r34,r35,r36,r37}. Specifically, the minimum cycle of a planar graph is the set of bounded faces\cite{r36,r37}.

\begin{figure}[htbp]
\begin{center}
\subfigure[]{
		\includegraphics[width=10cm]{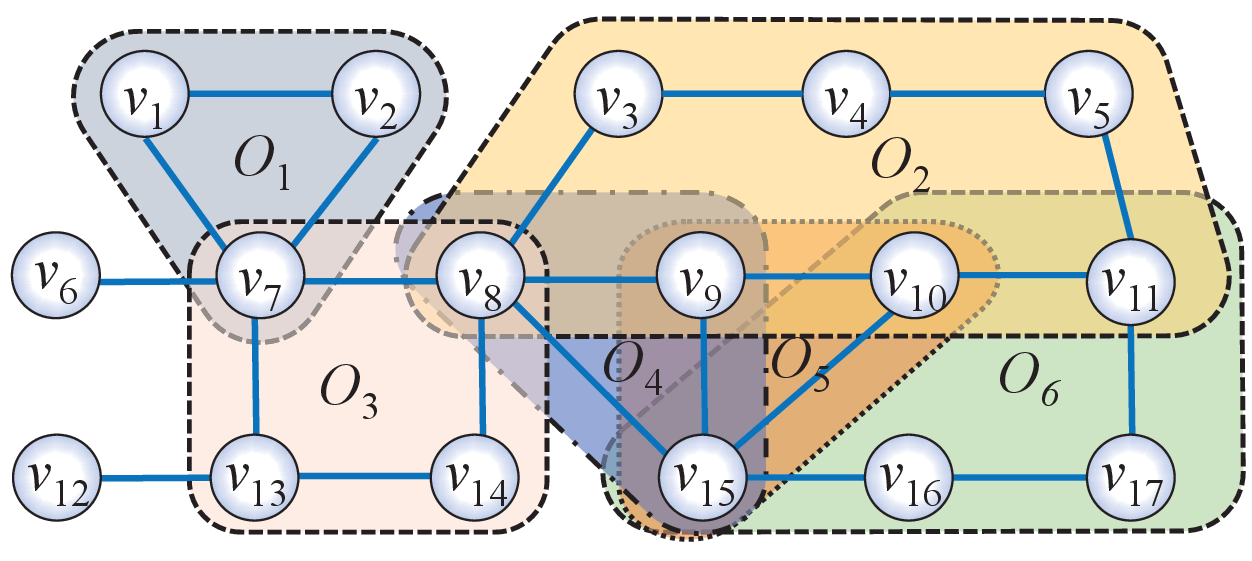}
		\label{fig1a} }

\subfigure[]{
		\includegraphics[width=3.8cm]{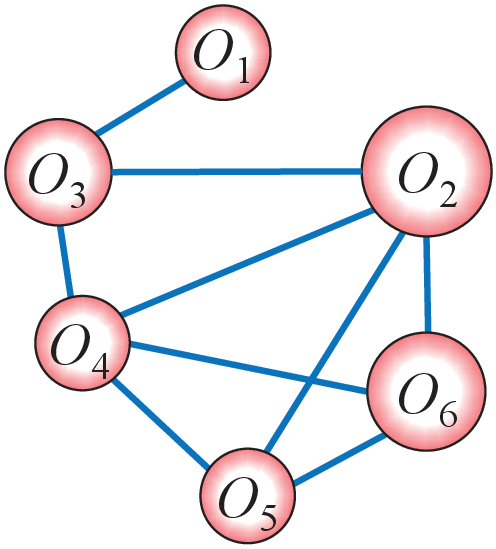}
		\label{fig1b} }\qquad \qquad
  \subfigure[]{
		\includegraphics[width=3.8cm]{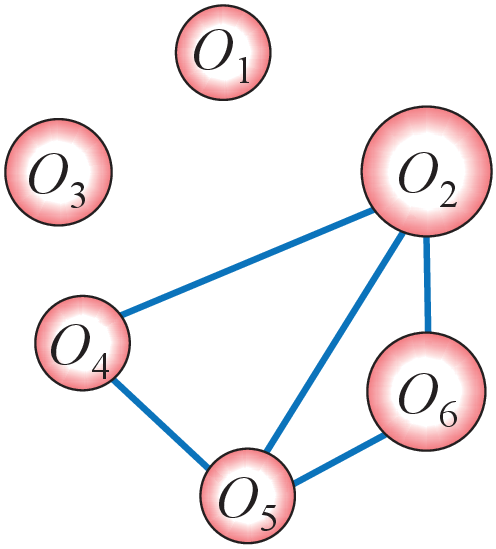}
		\label{fig1c} }

\caption{An example network and its cycle-based dual networks. (a)There are six shortest cycles: $O_{1}: v_{1}v_{2}v_{7}v_{1}$, $O_{2}: v_{3}v_{4}v_{5}v_{11}v_{10}v_{9}v_{8}v_{3}$, $O_{3}: v_{7}v_{8}v_{14}v_{13}v_{7}$, $O_{4}: v_{8}v_{9}v_{15}v_{8}$, $O_{5}: v_{9}v_{10}v_{15}v_{9}$ and $O_{6}: v_{10}v_{11}v_{17}v_{16}v_{15}v_{10}$. (b)The cycle-based dual network that establishes connections by sharing one node between the minimum cycle basis. (c)The cycle-based dual network that establishes connections by sharing one edge between the minimum cycle basis.}\label{fig1}
\end{center}
\end{figure}

For a node $v_{i}$, the degree is $k_i$, and the set of cycles associated with  $v_{i}$ is presented as $C_i = \{C_{i1}, C_{i2}, ~...,~C_{in_{i}}\}$, where $n_{i}$, called cycle degree, is the size of $C_i$. The length of each cycle in $C_i$ is denoted by $l_{i1}^{c}, l_{i2}^{c}, ~...,~l_{in_{i}}^{c}$. In order to measure the importance of nodes in the urban road network, we propose an observation based on the cycle, named cycle contribution rate,  which is defined as:
\begin{eqnarray}
\rho_{i} = \sum_{j=1}^{n_{i}}\frac{1}{l_{ij}^{c}}.
\label{e1}
\end{eqnarray}
In Ref.\cite{r40}, the cycle ratio is proposed as an effective measurement based on the shortest cycles to examine the importance of nodes. Similarly, the cycle ratio based on the minimum cycle basis is defined as:
\begin{eqnarray}
r_{i} = \sum_{j}^{n_{i}} \frac{n^{j}_{i}}{n_{j}},
\label{e2}
\end{eqnarray}
where $n^{j}_{i}$ represents the total number of cycles associated with nodes $v_{i}$ and $v_{j}$ simultaneously.
Fig.~\ref{fig1} is a schematic diagram of the basic form and cycle characteristics of a network. Table~\ref{Tab1} presents the degree, betweenness centrality, cycle degree, cycle ratio and cycle contribution rate of a node shown in Fig.~ \ref{fig1a}.

\begin{table}[htbp]
\centering
\caption{Node centrality of Fig.~\ref{fig1a}. $k$: degree; $BC$: betweeness centrality; $n$: cycle degree; $r$: cycle ratio; $\rho$: cycle contribute ratio.}
\label{Tab1}
\footnotesize
\begin{tabular}{@{}llccccc}
\br
\textbf{Node} & \textbf{Associated cycles} & $\bm{k}$ & $\bm{BC}$ & $\bm{n}$ & $\bm{r}$ & $\bm{\rho}$ \\
\mr
$v_{1}$ & $O_{1}$ & 2 & 0 & 1 & 2.5 & 0.33 \\
$v_{2}$ & $O_{1}$ & 2 & 0 & 1 & 2.5 & 0.33 \\
$v_{3}$ & $O_{2}$ & 2 & 18.50 & 1 & 4.5 & 0.14 \\
$v_{4}$ & $O_{2}$ & 2 & 10.50 & 1 & 4.5 & 0.14 \\
$v_{5}$ & $O_{2}$ & 2 & 5 & 1 & 4.5 & 0.14 \\
$v_{6}$ & Null & 1 & 0 & 0 & 0 & 0 \\
$v_{7}$ & $O_{1}$, $O_{3}$ & 5 & 51 & 2 & 5.33 & 0.58 \\
$v_{8}$ & $O_{2}$, $O_{3}$, $O_{4}$ & 5 & 72 & 3 & 8.33 & 0.73 \\
$v_{9}$ & $O_{2}$, $O_{4}$, $O_{5}$ & 3 & 8.05 & 3 & 6.50 & 0.81 \\
$v_{10}$ & $O_{2}$, $O_{5}$, $O_{6}$ & 3 & 12.50 & 3 & 8.67 & 0.68 \\
$v_{11}$ & $O_{2}$, $O_{6}$ & 3 & 9.50 & 2 & 7.67 & 0.35 \\
$v_{12}$ & Null & 1 & 0 & 0 & 0 & 0 \\
$v_{13}$ & $O_{3}$ & 3 & 17 & 1 & 2.83 & 0.25 \\
$v_{14}$ & $O_{3}$ & 2 & 10 & 1 & 2.83 & 0.25 \\
$v_{15}$ & $O_{4}$, $O_{5}$, $O_{6}$ & 4 & 29 & 3 & 5.17 & 0.87 \\
$v_{16}$ & $O_{6}$ & 2 & 10 & 1 & 3.17 & 0.20 \\
$v_{17}$ & $O_{6}$ & 2 & 2.50 & 1 & 3.17 & 0.20 \\
\br
\end{tabular}
\end{table}

For an undirected network $G$, define the set of cycles as $C^{G}=\bigcup_{i=1}^{N}C_{i} = \{ C_{1}^{G}, C_{1}^{G},~ ...~, C_{N^{o}}^{G}\} $, where $N^{o}$ represents the total number of the minimum cycle basis in the network $G$, and the corresponding the minimum cycle basis length is $l_{1}^{G},l_{2}^{G},~...~,l_{N^{o}}^{G}$, then the average length of the minimum cycle basis is defined as:
\begin{eqnarray}
\left \langle L^{c} \right \rangle = \frac{1}{N^{o}} \sum_{i=1}^{N^{o}} l_{i}^{G}.
\label{e3}
\end{eqnarray}

To investigate the overall properties of the network, we conducted a robustness analysis, which involves sequentially removing nodes from the network according to certain rules.
six strategies are employed to remove nodes from the network based on random, degree, betweenness, cycle degree, cycle ratio and cycle contribution rate.
The relative size of the maximum connected subgraph $S$, the average length $\left \langle L^{c} \right \rangle$ and the total number $ N^{o} $ of shortest cycles in the network as evaluation metrics are selected to analyze the network robustness. $S$ is denoted as:
\begin{eqnarray}
S = \frac{N'}{N},
\label{e5}
\end{eqnarray}
where $N$ represents the total number of nodes in the initial network, and $N'$ represents the total number of nodes in the maximum connected subgraph after node removal.

\subsection{Cycle-based dual network}
To analyze the interrelation among individual cycles, cycle-based dual network is constructed in this paper, each cycle is treated as a network node, and if two cycles share common nodes, a corresponding edge is established. Two cycle-based dual networks are constructed in Fig.~\ref{fig1b} and Fig.~\ref{fig1c} when each pair cycles shared one node and each pair cycles shared one edge in Fig.~\ref{fig1a}.

\section{Results and discussions}\label{S3}
\subsection{Data}
\begin{figure}
	\centering
	\includegraphics[width=15cm]{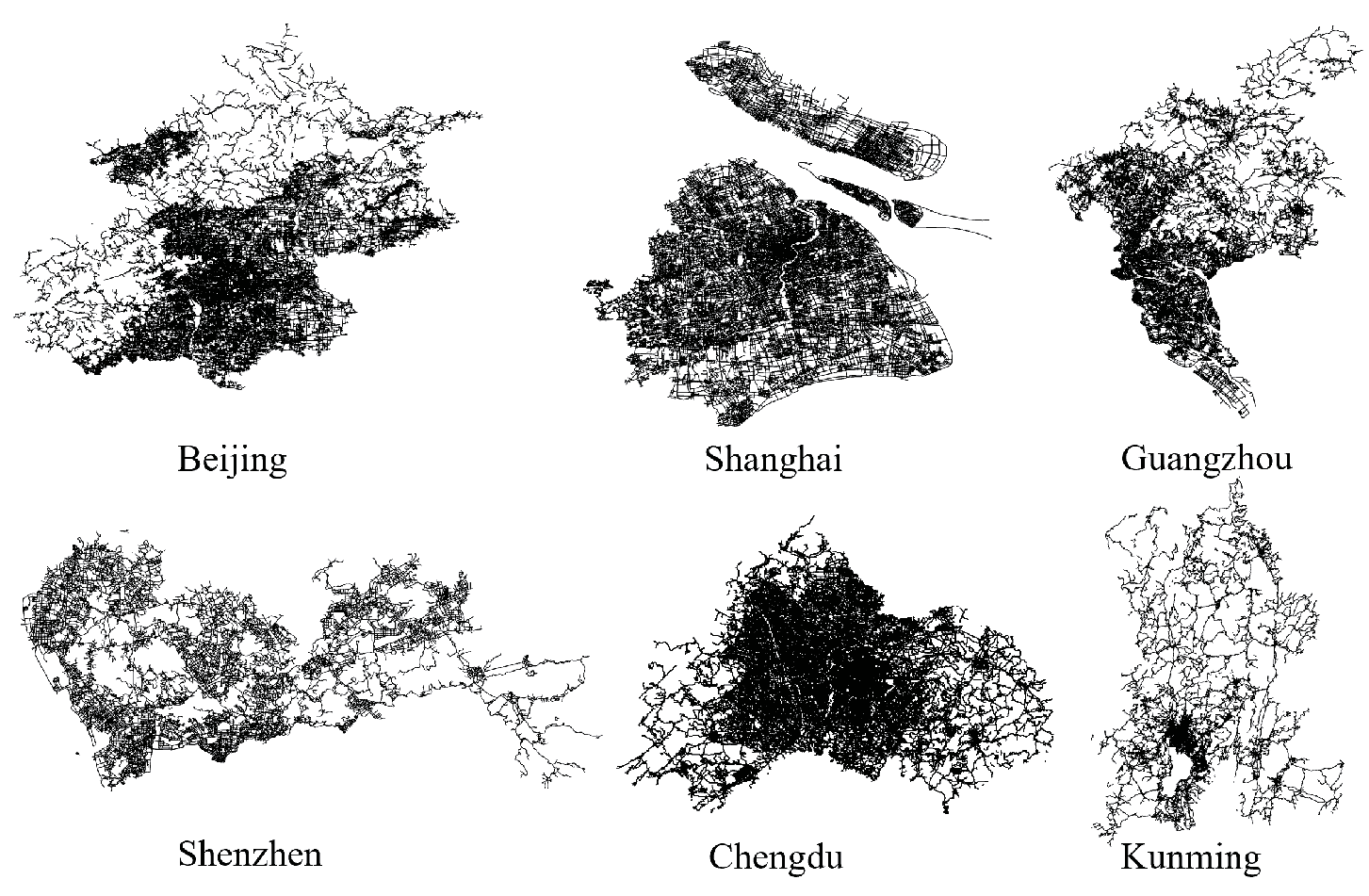}
\caption{Experimental urban road networks.}\label{fig2}
\end{figure}

Population, economy and comprehensive strength are the key indicators to evaluate urban development. This study focuses on six cities in China(Beijing, Shanghai, Guangzhou, Shenzhen, Chengdu and Kunming) to explore the potential information of urban road networks. Fig.~\ref{fig2} visualizes the geographical morphology of the six road networks.

The urban road network data used in this study is obtained from the official OpenStreetMap (OSM) website (www.openstreetmap.org). The OSM database stores a huge environmental spatial dataset with very rich metadata covering the major cities of the world. The OSM network data of 6 cities in China were obtained on September 20, 2023 in shp vector data format and the projection coordinate system was WGS\_1984\_Mercator coordinate system.

OSM road data has good integrity, but the initial data is rough. In order to improve the quality of road network data and the accuracy of experimental results, data preprocessing is essential. In this paper, data preprocessing includes verifying data consistency, eliminating overlapping paths and invalid data, and then converting double lines into single lines, and finally converting single lines into simple graph structure. In addition, the topology of the network needs to be checked and isolated roads removed to ensure network connectivity.

\begin{figure}
	\centering
	\includegraphics[width=14cm]{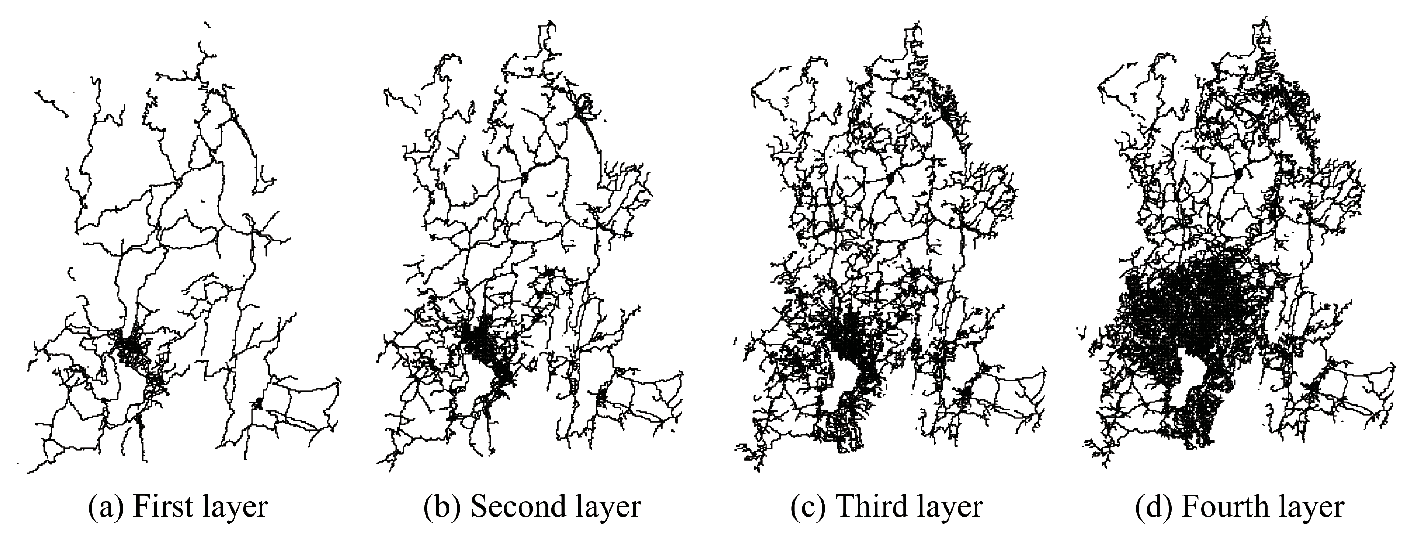}
\caption{Different layers of urban road networks in Kunming.}\label{fig3}
\end{figure}

\begin{figure}
	\centering
	\includegraphics[width=14cm]{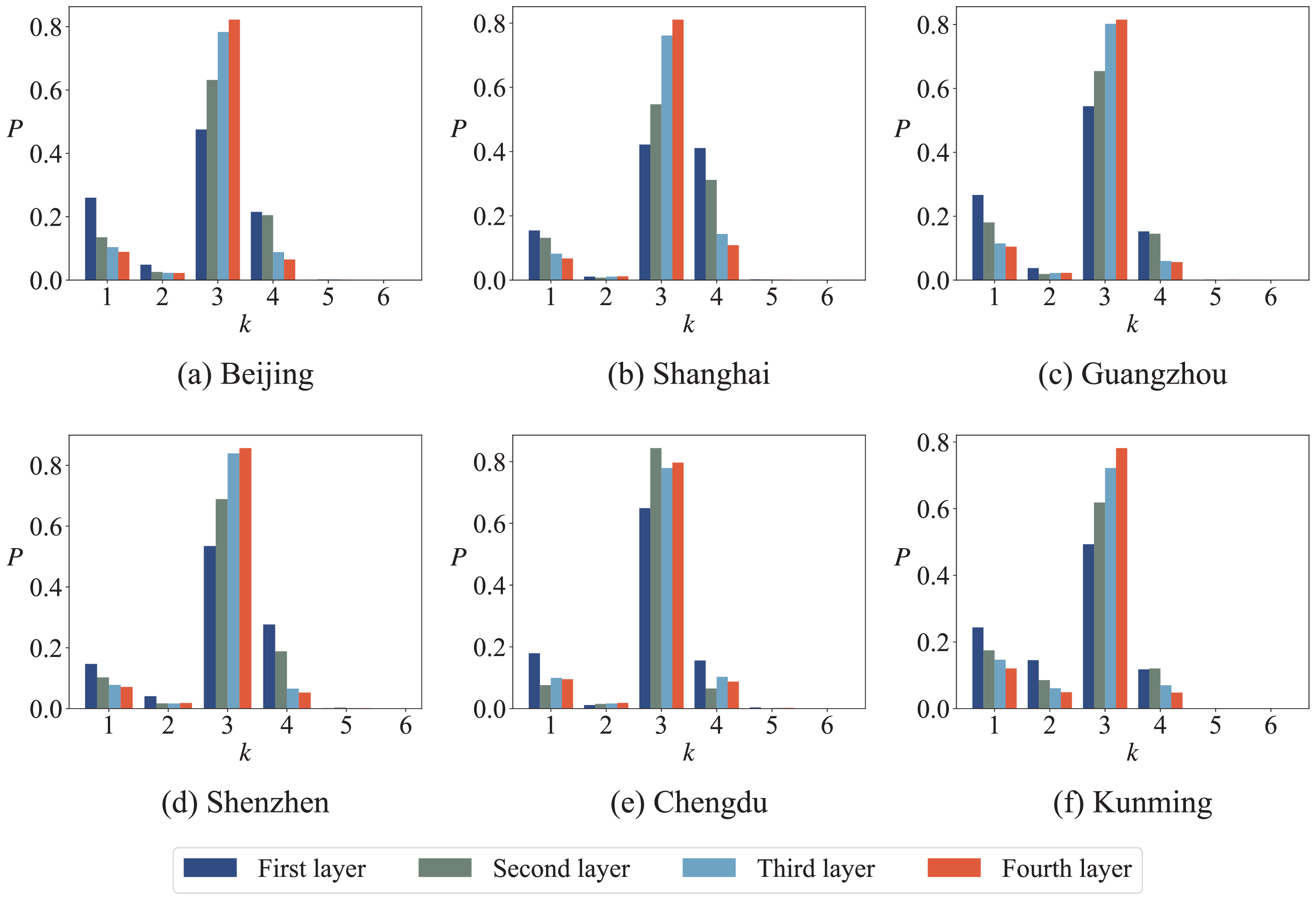}
\caption{The degree distribution of urban road network at different layers. In all plots, the $x$-axis label $k$ represents the degree of nodes and the $y$-axis label $P$ represents the proportion of each part.}\label{fig4}
\end{figure}

We choose four road categories to model from OSM road data, which include main roads, secondary trunk roads, branch roads and internal roads. The main road is the backbone of the urban road network, which includes primary and secondary field in OSM. The Secondary trunk road is a regional traffic trunk road in the city, which includes tertiary field in OSM. The branch road is the connection line of the secondary trunk road connecting each residential area, which includes residential and unclassified field in OSM. The internal road includes bridleway, living\_street, path, and service field in OSM.

In the modeling process, we divide the road network into four levels, namely the first layer network, the second layer network, the third layer network, and the fourth layer network. The first layer network only considers main roads, the second layer network adds secondary trunk roads on the basis of the first layer network, the third layer network introduces branch roads on the basis of the second layer network, and the fourth layer network adds internal roads on the basis of the third layer network. Fig.~\ref{fig3} shows the spatial distribution of the four layers road network.

Fig.~\ref{fig4} shows the degree distribution of urban road network at different layers respectively. It can be seen that the degree distribution of urban road network is discrete, only a few values can be obtained. In each layer of the network, nodes with degrees of 3 occupy a larger proportion, followed by nodes with degrees of 4 and 1, that is, T-shaped roads account for a larger proportion in the network, and there are also some intersections and damaged roads. It is well-known that degree distribution plays a significant role in various network fields, such as vital nodes identification, community detection, and link prediction. However, the degree distribution of roadway networks contains relatively limited information, which is unfavorable for the analysis and modeling of issues related to roadway networks. Therefore, in the following, we will attempt to analyze roadway networks using motifs and minimum cycle bases.

The topological characteristics of different layers in urban road network are showed in Table~\ref{Tab2}, which include the number of nodes, the number of edges, average degree, average path length, clustering coefficient and network diameter. We find that the average path length and network diameter in the urban road network are larger, the clustering coefficient is smaller.

\begin{table}[htbp]
\centering
\caption{The topological characteristics of urban road network. $N$: the number of nodes; $M$: the number of edges; $\langle k \rangle$: average degree; $\langle d \rangle$: average path length; $\langle C \rangle$: clustering coefficient; $D$: network diameter.}
\label{Tab2}
\footnotesize
\begin{tabular}{@{}llccccccc}
\br
\textbf{City} & \textbf{Level} & $\bm{N}$ & $\bm{M}$ & $\bm{\langle k \rangle}$ & $\bm{\langle d \rangle}$ & $\bm{\langle C \rangle}$ & $\bm{D}$ \\
\mr
\multirow{4}{*}{Beijing} & First layer & 3348 & 4432 & 2.65 & 34.97 & 0.05 & 106 \\
 & Second layer & 14294 & 20823 & 2.91 & 56.82 & 0.05 & 164 \\
 & Third layer & 65660 & 93949 & 2.86 & 138.59 & 0.05 & 515 \\
 & Fourth layer & 102930 & 147576 & 2.87 & 165.49 & 0.06 & 466 \\
  \mr
\multirow{4}{*}{Shanghai} & First layer & 2110 & 3265 & 3.09 & 23.23 & 0.04 & 62 \\
 & Second layer & 9827 & 14960 & 3.04 & 45.02 & 0.04 & 113 \\
 & Third layer & 48723 & 72409 & 2.97 & 91.58 & 0.04 & 229 \\
 & Fourth layer & 64201 & 95168 & 2.96 & 107.73 & 0.04 & 266 \\
  \mr
\multirow{4}{*}{Guangzhou} & First layer & 958 & 1237 & 2.58 & 17.48 & 0.07 & 40 \\
 & Second layer & 5096 & 7053 & 2.77 & 38.24 & 0.06 & 88 \\
 & Third layer & 31796 & 44679 & 2.81 & 92.39 & 0.06 & 245 \\
 & Fourth layer & 38416 & 54295 & 2.83 & 97.47 & 0.06 & 271 \\
  \mr
\multirow{4}{*}{Shenzhen} & First layer & 1424 & 2097 & 2.95 & 24.46 & 0.06 & 74 \\
 & Second layer & 5241 & 7794 & 2.97 & 48.44 & 0.05 & 129 \\
 & Third layer & 18485 & 26767 & 2.90 & 96.53 & 0.05 & 263 \\
 & Fourth layer & 23906 & 34605 & 2.90 & 111.56 & 0.05 & 313 \\
  \mr
\multirow{4}{*}{Chengdu} & First layer & 2372 & 3316 & 2.80 & 28.50 & 0.04 & 84 \\
 & Second layer & 10280 & 14907 & 2.90 & 51.09 & 0.05 & 124 \\
 & Third layer & 58947 & 85251 & 2.89 & 118.49 & 0.05 & 320 \\
 & Fourth layer & 64966 & 93662 & 2.88 & 123.51 & 0.05 & 331 \\
 \mr
\multirow{4}{*}{Kunming} & First layer & 1310 & 1629 & 2.49 & 36.52 & 0.07 & 85 \\
 & Second layer & 4634 & 6227 & 2.69 & 57.21 & 0.08 & 165 \\
 & Third layer & 17231 & 23411 & 2.72  & 94.87 & 0.07 & 260 \\
 & Fourth layer & 33623 & 46411 & 2.76 & 115.12 & 0.08 & 313 \\
\br
\end{tabular}
\end{table}

\subsection{Motif}
\begin{figure}
	\centering
	\includegraphics[width=14cm]{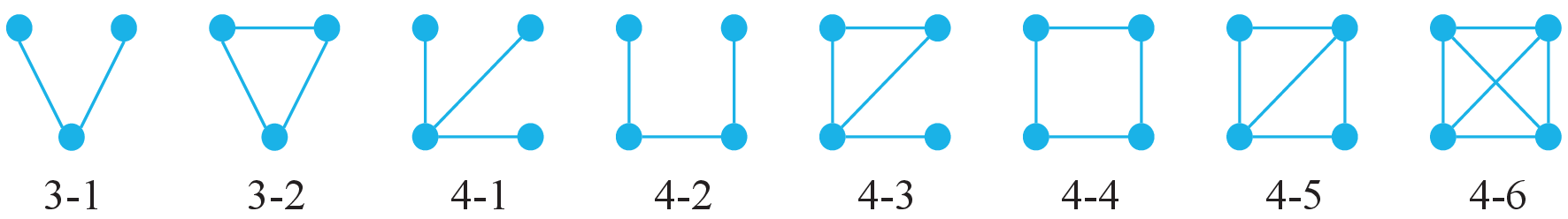}
\caption{The schematic diagram of motifs consisting of three or four nodes in undirected network. }\label{fig5}
\end{figure}

Motif is a fundamental pattern of repeated interactions in networks that occur significantly more frequently than random networks with the same number of nodes and edges. Eight motifs of undirected network are showed in Fig.~\ref{fig5}, including two three-node motifs and six four-node motifs.

To measure the importance of motifs in real-world networks, it is necessary to generate a large number of random networks that are consistent with the degree distribution of real-world networks. To compare the frequency of motifs in real-world networks with those in random networks, the Z-score is introduced. For a motif $i$, the Z-score is calculated as\cite{r47}:
\begin{eqnarray}
Z_{i} = \frac{N_{real_{i}} - \left \langle N_{rand_{i}} \right \rangle }{\sigma_{rand_{i}} },
\label{e1}
\end{eqnarray}
where $N_{real_{i}}$ is the frequency of occurrence of motif $i$ in real-world network, $\left \langle N_{rand_{i}} \right \rangle$ and  $\sigma_{rand_{i}}$ are obtained from 1000 different random networks with the same degree distribution of real-world network, representing the average frequency and standard deviation respectively.

From Eq.~\ref{e1}, when $N_{real_{i}} > \left \langle N_{rand_{i}} \right \rangle$, the Z-score is positive, which indicates that the motif occurs more frequently in the real-world network. when $N_{real_{i}} < \left \langle N_{rand_{i}} \right \rangle$, the Z-score is negative, which indicates that the motif occurs less frequently in the real-world network.

\begin{table}[htbp]
\centering
\caption{Z-score for different layers of urban road network.}
\label{Tab3}
\footnotesize
\begin{tabular}{@{}llcccccccc}
\br
\textbf{City} & \textbf{Level} & \textbf{3-1} & \textbf{3-2} & \textbf{4-1} & \textbf{4-2} & \textbf{4-3} & \textbf{4-4} & \textbf{4-5} & \textbf{4-6} \\
\mr
\multirow{4}{*}{Beijing} & First layer & -141.70  & 141.70  & -118.07  & -63.99  & 114.26  & 305.99  & $\infty$ & - \\
& Second layer & -579.03  & 579.03  & -469.11  & -240.63  & 451.07  & 1456.05  & $\infty$ & - \\
& Third layer & -3047.24  & 3047.24  & -2813.78  & -756.64  & 2732.62  & 5268.11  & $\infty$ & - \\
& Fourth layer & -4526.36  & 4526.36  & -4086.01  & -1226.12  & 3970.74  & 7789.24  & $\infty$ & - \\
\mr
\multirow{4}{*}{Shanghai} & First layer & -78.68  & 78.68  & -66.21  & -105.91  & 64.63  & 260.17  & $\infty$ & - \\
& Second layer & -253.55 & 253.55 & -228.89 & -206.16 & 224.78 & 1162.15 & $\infty$ & - \\
& Third layer & -1782.06  & 1782.06  & -1609.71  & -611.13  & 1571.48  & 5186.93  & $\infty$ & -  \\
& Fourth layer & -2083.12  & 2083.12  & -1937.58  & -924.20  & 1897.10  & 5609.74  & $\infty$ & - \\
\mr
\multirow{4}{*}{Guangzhou} & First layer & -54.12  & 54.12  & -46.51  & -36.29  & 46.02  & 62.40  & $\infty$ & - \\
& Second layer & -268.79  & 268.79  & -245.67  & -128.32  & 237.82  & 447.30  & $\infty$ & - \\
& Third layer & -1666.81  & 1666.81  & -1597.69  & -578.39  & 1557.39  & 2611.52  & $\infty$ & - \\
& Fourth layer & -1779.24  & 1779.24  & -1718.14  & -659.79  & 1671.67  & 2828.27  & $\infty$ & - \\
\mr
\multirow{4}{*}{Shenzhen} & First layer & -67.63  & 67.63  & -56.45  & -77.69  & 53.23  & 169.11  & $\infty$ & - \\
& Second layer & -253.92  & 253.92  & -224.53  & -164.04  & 217.47  & 504.77  & $\infty$ & - \\
& Third layer & -895.10  & 895.10  & -821.19  & -478.17  & 792.83  & 1735.01  & $\infty$ & - \\
& Fourth layer & -1042.45  & 1042.45  & -973.23  & -692.74  & 941.60  & 2052.26  & $\infty$ & - \\
\mr
\multirow{4}{*}{Chengdu} & First layer & -84.39  & 84.39  & -79.59  & -70.49  & 78.08  & 211.80  & $\infty$ & - \\
& Second layer & -364.80  & 364.80  & -334.94  & -341.94  & 326.87  & 917.95  & $\infty$ & - \\
& Third layer & -2500.57  & 2500.57  & -2238.12  & -760.30  & 2183.14  & 5054.40  & $\infty$ & - \\
& Fourth layer & -2807.33  & 2807.33  & -2521.32  & -852.74  & 2457.29  & 5838.93  & $\infty$ & - \\
\mr
\multirow{4}{*}{Kunming} & First layer & -78.17  & 78.17  & -68.40  & -40.97  & 63.27  & 91.25  & $\infty$ & - \\
& Second layer & -275.56  & 275.56  & -241.95  & -118.20  & 227.13  & 350.02  & $\infty$ & - \\
& Third layer & -1199.16  & 1199.16  & -1064.96  & -329.96  & 1009.31  & 1445.67  & $\infty$ & - \\
& Fourth layer & -1957.88  & 1957.88  & -1854.06  & -672.93  & 1768.70  & 3073.76  & $\infty$ & - \\
\br
\end{tabular}
\end{table}

As shown in Table~\ref{Tab3}, Z-scores of each motif in different layers road network of Beijing, Shanghai, Guangzhou, Shenzhen, Chengdu and Kunming are summarized. It can be observed that the Z-score of motifs with cyclic structure, such as the Z-scores of 3-2, 4-3 and 4-4 are positive, and their occurrence frequency are significantly higher than that of random networks with sequences of the same degree, while the Z-scores of 3-1, 4-1 and 4-2 are negative. In particular, 4-5 hardly appears in random networks and 4-6 never appears both in road networks and in random networks. That is to say, the frequency of motifs that contain cycles surpasses that of random networks with equivalent degree sequences. Thus, we conduct topological and robustness analyses of urban road networks using the minimum cycle basis as fundamental components in the following sections.

\subsection{Urban road analysis based on the minimum cycle basis}
The topological characteristics based on cycle of different layers in urban road network are showed in Table~\ref{Tab4}, which include the number of nodes, the number of edges, average degree, the total number of cycles, the average length of the minimum cycle basis, the ratio of nodes in the cycles to total nodes, the ratio of edges in the cycles to total edges. We can see that the values of $N^{c}/N$ and $M^{c}/M$ are around 80\%. That means the majority of nodes and edges are associated with the minimum cycle basis, thus the study of urban road based on the minimum cycle basis is meaningful.
\begin{table}[htbp]
\centering
\caption{The topological characteristics associated with cycles of the urban road network. $N^{o}$: the total number of cycles; $ \langle L^{c} \rangle$: the average length of the minimum cycle basis; $N^{c}$: the number of nodes within cycles; $M^{c}$: the number of edges within cycles; $N^{c}/N$: the ratio of nodes in the cycles to total nodes; $M^{c}/M$: the ratio of edges in the cycles to total edges.}
\label{Tab4}
\footnotesize
\begin{tabular}{@{}llccccccc}
\br
\textbf{City} & \textbf{Level} & $\bm{N}$ & $\bm{M}$ & $\bm{\langle k \rangle} $ & $\bm{N^{o}}$ & $\bm{\langle L^{c} \rangle}$ & $\bm{N^{c}/N}$ & $\bm{M^{c}/M}$ \\
\mr
\multirow{4}{*}{Beijing} & First layer & 3348 & 4432 & 2.65 & 1167 & 5.23 & 0.69 & 0.77 \\
& Second layer & 14294 & 20823 & 2.91 & 6656 & 5.32 & 0.84 & 0.89 \\
& Third layer & 65660 & 93949 & 2.86 & 27170 & 5.87 & 0.81 & 0.85 \\
& Fourth layer & 102930 & 147576 & 2.87 & 42887 & 5.93 & 0.84 & 0.87 \\
\mr
\multirow{4}{*}{Shanghai} & First layer & 2110 & 3265 & 3.09 & 1175 & 4.80 & 0.84 & 0.90 \\
& Second layer & 9827 & 14960 & 3.04 & 5173 & 5.10 & 0.86 & 0.91 \\
& Third layer & 48723 & 72409 & 2.97 & 23022 & 5.77 & 0.87 & 0.90 \\
& Fourth layer & 64201 & 95168 & 2.96 & 29936 & 5.85 & 0.88 & 0.91 \\
\mr
\multirow{4}{*}{Guangzhou} & First layer & 958 & 1237 & 2.58 & 305 & 5.12 & 0.66 & 0.75 \\
& Second layer & 5096 & 7053 & 2.77 & 2029 & 5.39 & 0.79 & 0.85 \\
& Third layer & 31796 & 44679 & 2.81 & 13019 & 5.83 & 0.87 & 0.90 \\
& Fourth layer & 38416 & 54295 & 2.83 & 16029 & 5.87 & 0.88 & 0.91 \\
\mr
\multirow{4}{*}{Shenzhen} & First layer & 1424 & 2097 & 2.95 & 684 & 4.99 & 0.83 & 0.89 \\
& Second layer & 5241 & 7794 & 2.97 & 2569 & 5.32 & 0.89 & 0.93 \\
& Third layer & 18485 & 26767 & 2.90 & 8312 & 5.79 & 0.91 & 0.94 \\
& Fourth layer & 23906 & 34605 & 2.90 & 10753 & 5.84 & 0.92 & 0.95 \\
\mr
\multirow{4}{*}{Chengdu} & First layer & 2372 & 3316 & 2.80 & 979 & 5.40 & 0.80 & 0.86 \\
& Second layer & 10280 & 14907 & 2.90 & 4658 & 5.83 & 0.92 & 0.94 \\
& Third layer & 58947 & 85251 & 2.89 & 26618 & 5.91 & 0.88 & 0.91 \\
& Fourth layer & 64966 & 93662 & 2.88 & 28755 & 5.97 & 0.87 & 0.91 \\
\mr
\multirow{4}{*}{Kunming} & First layer & 1310 & 1629 & 2.49 & 351 & 5.14 & 0.62 & 0.69 \\
& Second layer & 4634 & 6227 & 2.69 & 1642 & 5.38 & 0.75 & 0.81 \\
& Third layer & 17231 & 23411 & 2.72 & 6340 & 5.84 & 0.81 & 0.86 \\
& Fourth layer & 33623 & 46411 & 2.76 & 13051 & 5.98 & 0.85 & 0.89 \\
\br
\end{tabular}
\end{table}

\subsubsection{The length and surface areas distribution of the minimum cycle base}
\begin{figure}
	\centering
	\includegraphics[width=14cm]{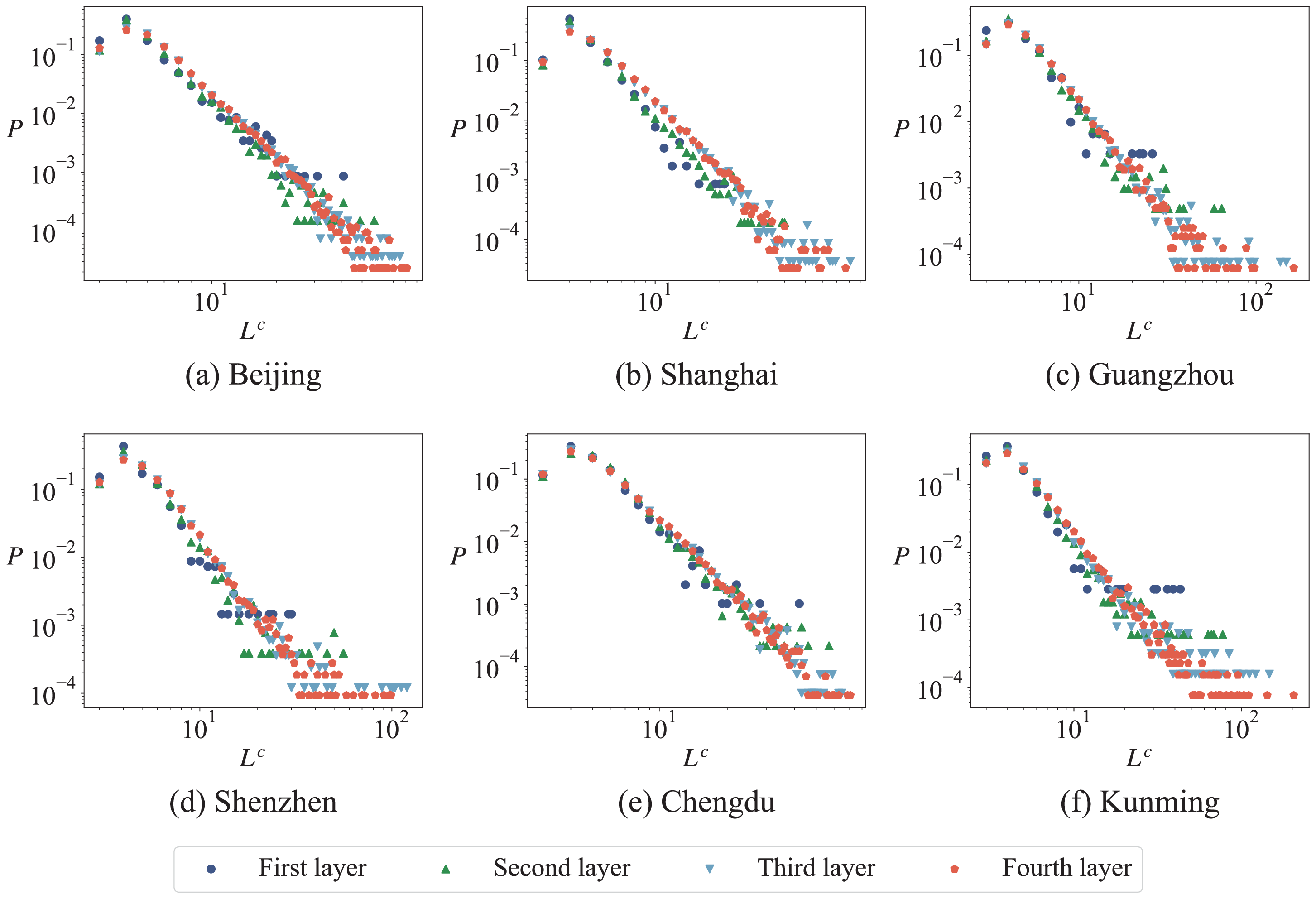}
\caption{The length distribution of the minimum cycle base in the urban road network for different cities. In all plots, the $x$-axis label $L^{c}$ represents the length of the minimum cycle basis and the $y$-axis label $P$ represents the proportion of each part.}
\label{fig6}
\end{figure}

\begin{figure}
	\centering
	\includegraphics[width=14cm]{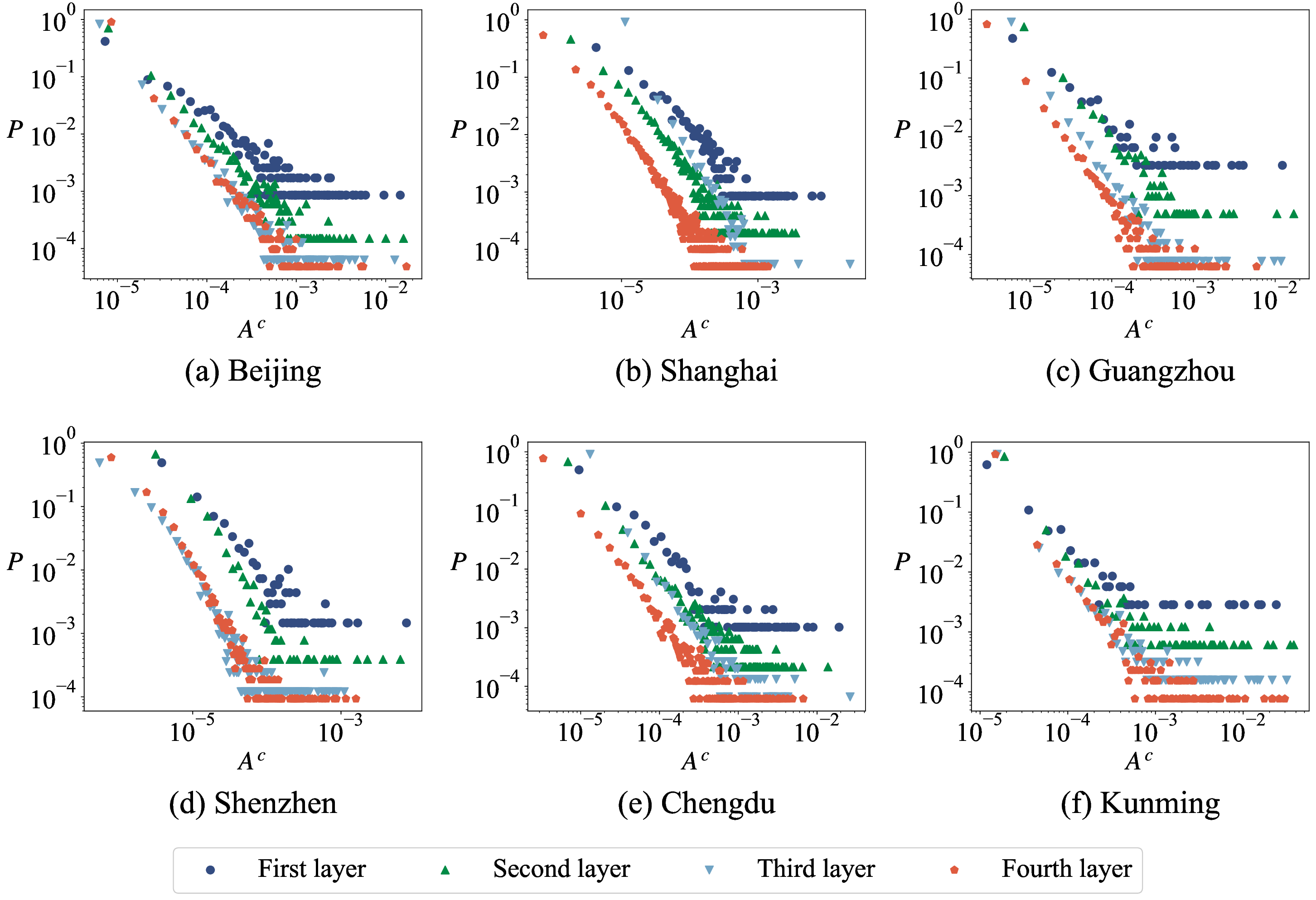}
\caption{The surface areas distribution of the minimum cycle base in the urban road network for different cities. In all plots, the $x$-axis label $A^{c}$ represents the surface areas and the $y$-axis label $P$ represents the proportion of each part.}
\label{fig7}
\end{figure}

Previous studies have confirmed that the frequency distribution of length and surface areas obey the power-law distribution when the cellular structures or called closed polygons are used to study road networks\cite{r38,r39,r40}. Fig.~\ref{fig6} and Fig.~\ref{fig7} show the length and surface areas distribution of the minimum cycle basis follow power-law distribution, with the exception of cycles comprised of three nodes. In other words, in the network, the minimum cycle basis with smaller length appears more frequently than that with larger length. Table~\ref{Tab5} shows the power index of the distribution of the cycle length of the road network at different layers in six cities, and it can be seen that the power indexes are different in different cities. During the modeling and analysis of urban road networks based on the minimum cycle basis, it is observed that the distribution of lengths and surface areas within the minimum cycle basis does not adhere to uniformity or normality; instead, it follows a power-law distribution. The modeling conducted previously using Voronoi tessellation evidently fails to satisfy this requirement\cite{Li2017}.

\begin{table}[htbp]
\centering
\caption{The power-law exponent of the length distribution.}
\footnotesize
\label{Tab5}
\begin{tabular}{@{}lcccc}
\br
\textbf{City} &  \textbf{First layer} &  \textbf{Second layer} &  \textbf{Third layer} &  \textbf{Fourth layer} \\
\mr
Beijing	& 3.43 & 4.08 & 3.62 & 3.64  \\
Shanghai & 4.15 & 3.88 & 3.76 & 3.83  \\
Guangzhou & 3.69 & 3.65 & 3.55 & 3.52  \\
Shenzhen & 3.78 & 3.96 & 3.85 & 3.78  \\
Chengdu	& 4.14 & 3.86 & 3.59 & 3.56  \\
Kunming	& 3.53 & 3.35 & 3.10 & 3.12  \\
\br
\end{tabular}
\end{table}

\subsubsection{Vital nodes identification based on the minimum cycle basis}
\begin{figure}
	\centering
	\includegraphics[width=9.3cm]{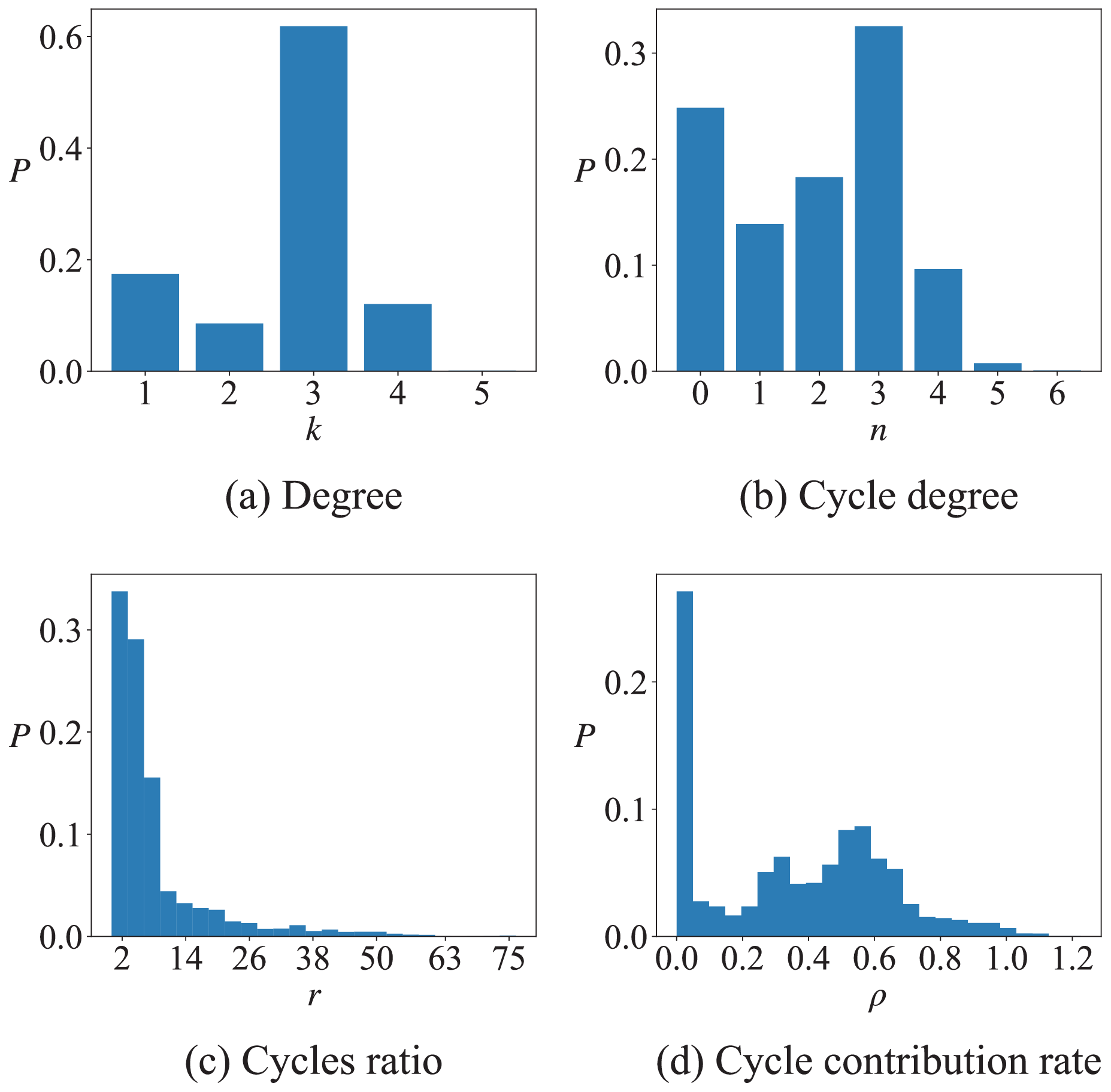}
\caption{The histogram of four node centrality measurements at the second layer of Kunming road network. (a) Degree; (b) Cycle degree; (c) Cycle ratio; (d) Cycle contribution rate.}\label{fig8}
\end{figure}

Taking the second layer of Kunming road network as an example, the value distribution of degree, cycle degree, cycle ratio and cycle contribution rate are also showed in Fig.~\ref{fig8}. The values of degree and cycle degree are only a few discrete numbers, while the values of cycle ratio and cycle contribution rate have a lot of possibilities. The distribution of cycle ratio exhibits power-law characteristics, while the cycle contribution rate demonstrates normal distribution characteristics. In the context of node ranking, the cycle contribution rate exhibits superior discriminatory power among nodes in comparison to the cycle ratio.

\begin{figure}[htbp]
\begin{center}
\subfigure[Degree]{
		\includegraphics[width=6cm]{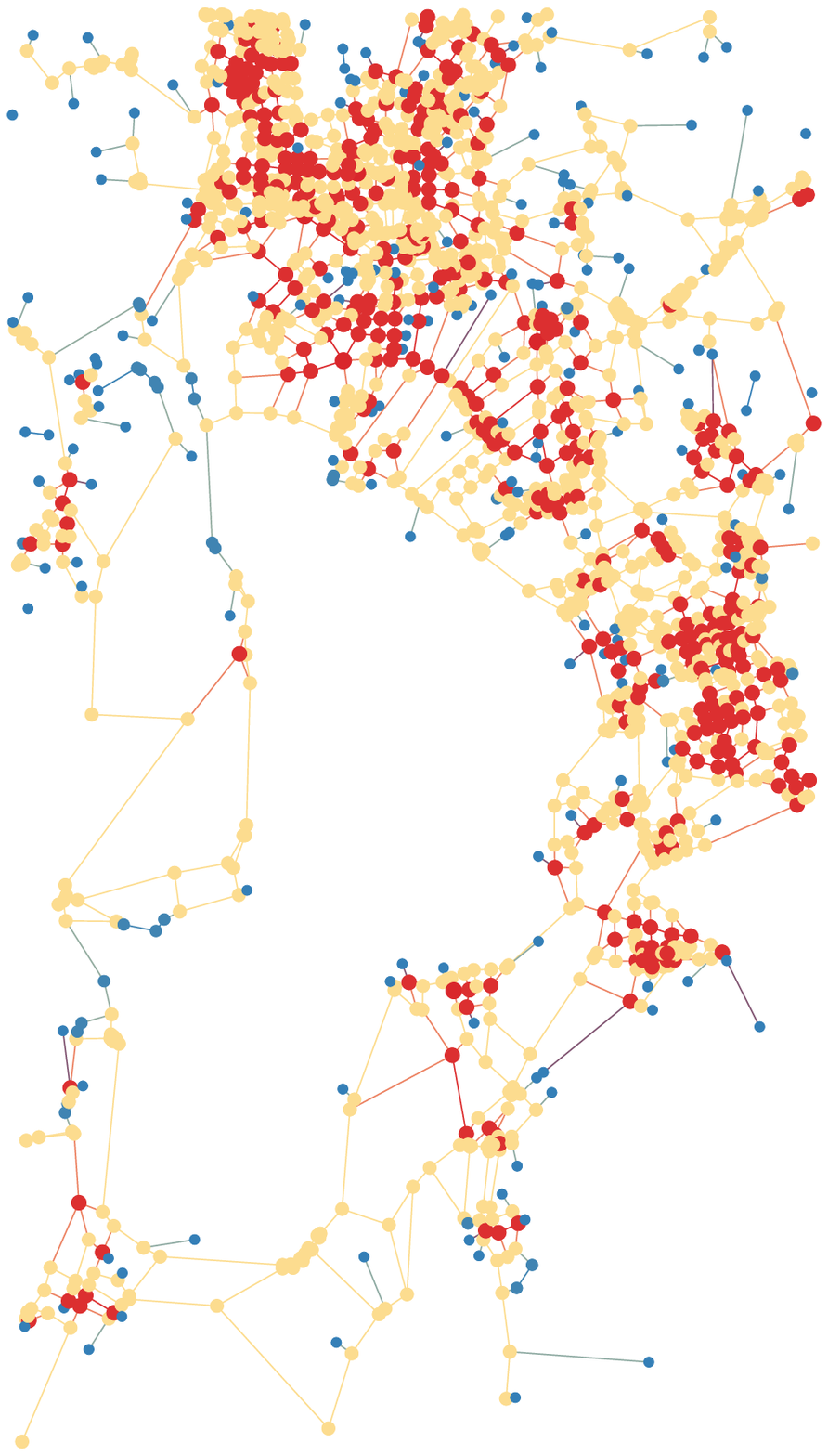}
		\label{fig9a}  }
\subfigure[Cycle degree]{
		\includegraphics[width=6cm]{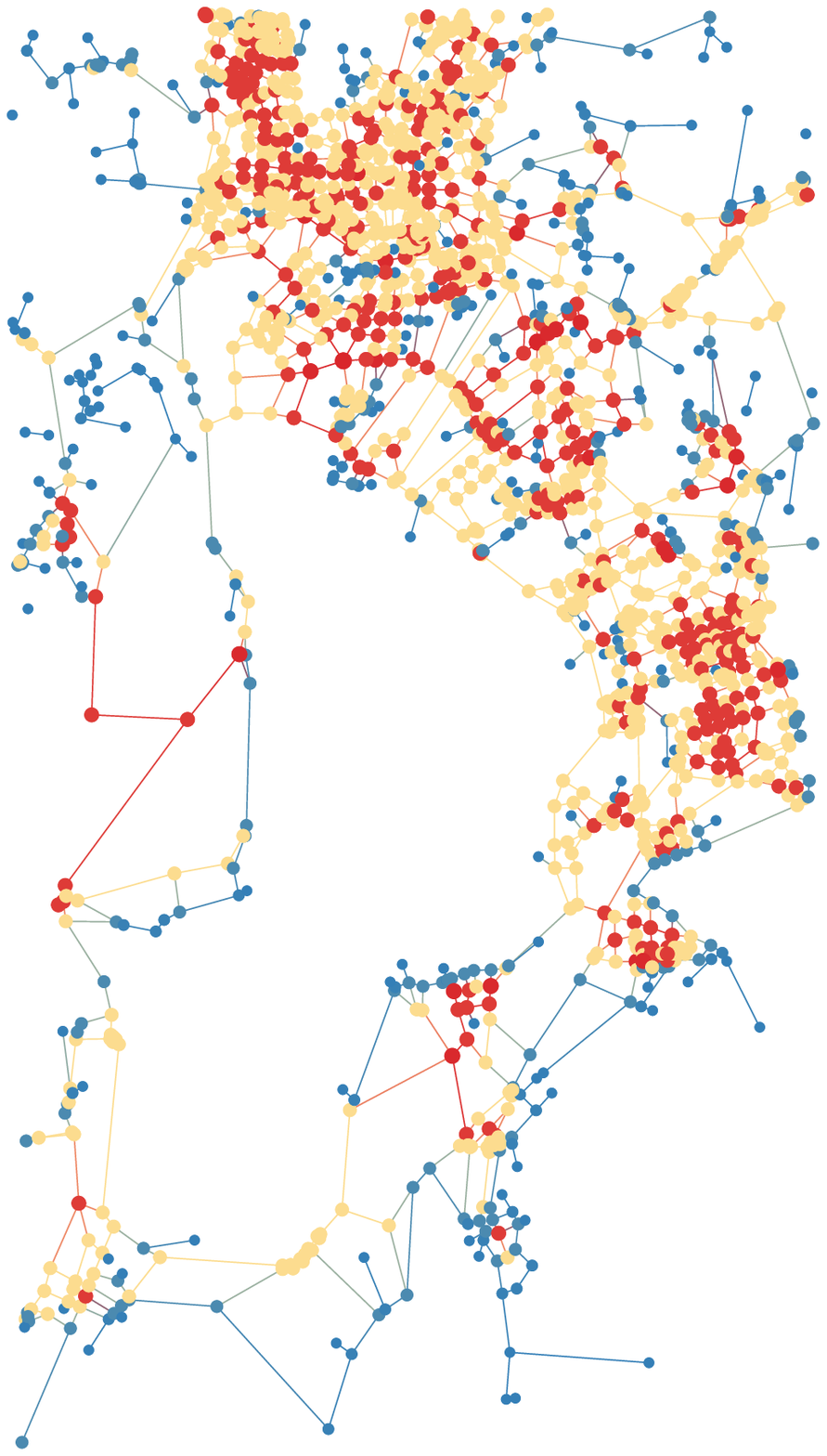}
		\label{fig9b} }
\subfigure[Cycle ratio]{
		\includegraphics[width=6cm]{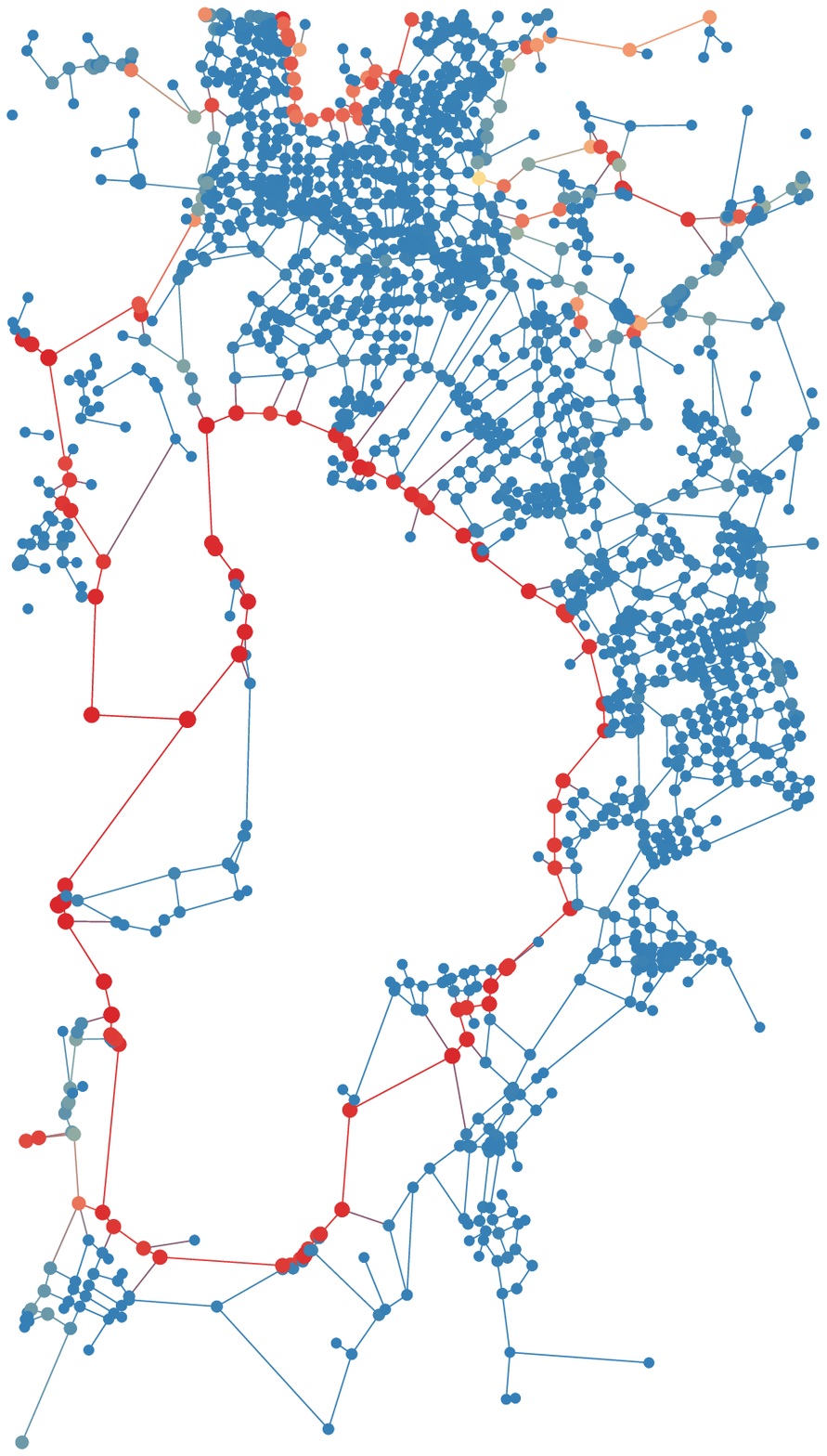}
		\label{fig9c} }
\subfigure[Cycle contribution rate]{
		\includegraphics[width=6cm]{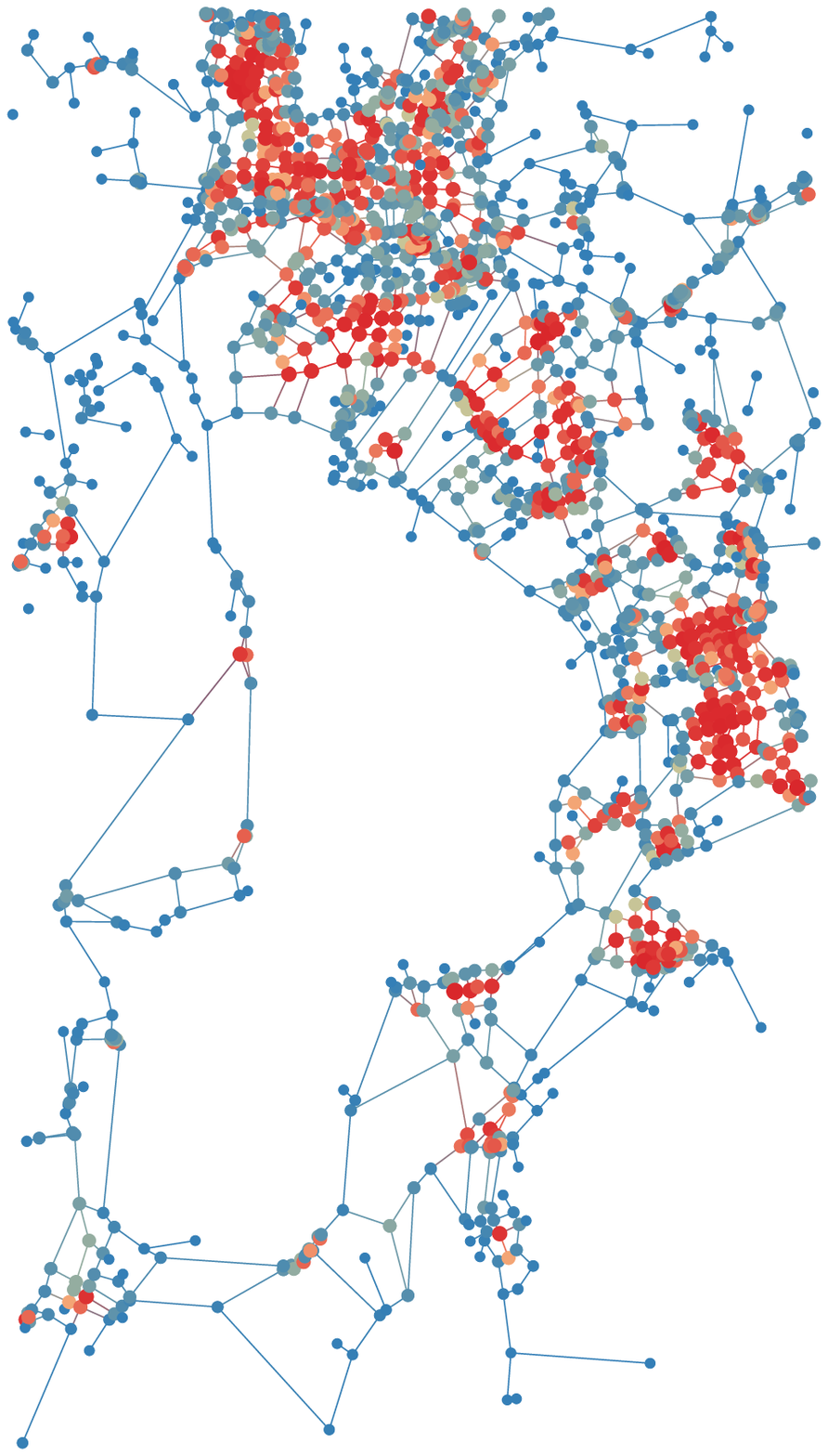}
		\label{fig9d}  }
\subfigure{
		\includegraphics[width=4.5cm]{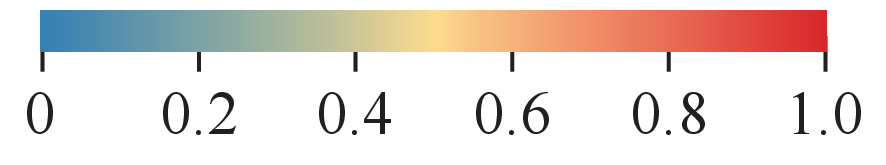}
		\label{fig9e}  }
\caption{Localized spatial distribution for different normalized node centrality measurements, including degree, cycle degree, cycle ratio and cycle contribution rate, at the second layer of Kunming road network.}
\label{fig9}
\end{center}
\end{figure}

The spatial distribution of degree, cycle degree, cycle ratio and cycle contribution rate are showed in Fig.~\ref{fig9}. We can see that there are almost three  kinds of node with different colors, blue (indicates small values), yellow (indicates medium values) and red (indicates large values) in Fig.~\ref{fig9a} and Fig.~\ref{fig9b}. Obviously, the nodes are hard to distinguish by degree and cycle degree. Nodes possessing a high cycle ratio are typically situated within larger cycles in Fig.~\ref{fig9c}. As shown in Fig.~\ref{fig9d}, nodes with higher cycle contribution rates are typically located in the core area of the network. This result can be used to identify the polycentricity of the network, and this discovery will aid in the selection and identification of urban material hubs and commercial centers.

\subsubsection{Robustness Analysis}
\begin{figure}
	\centering
	\includegraphics[width=14cm]{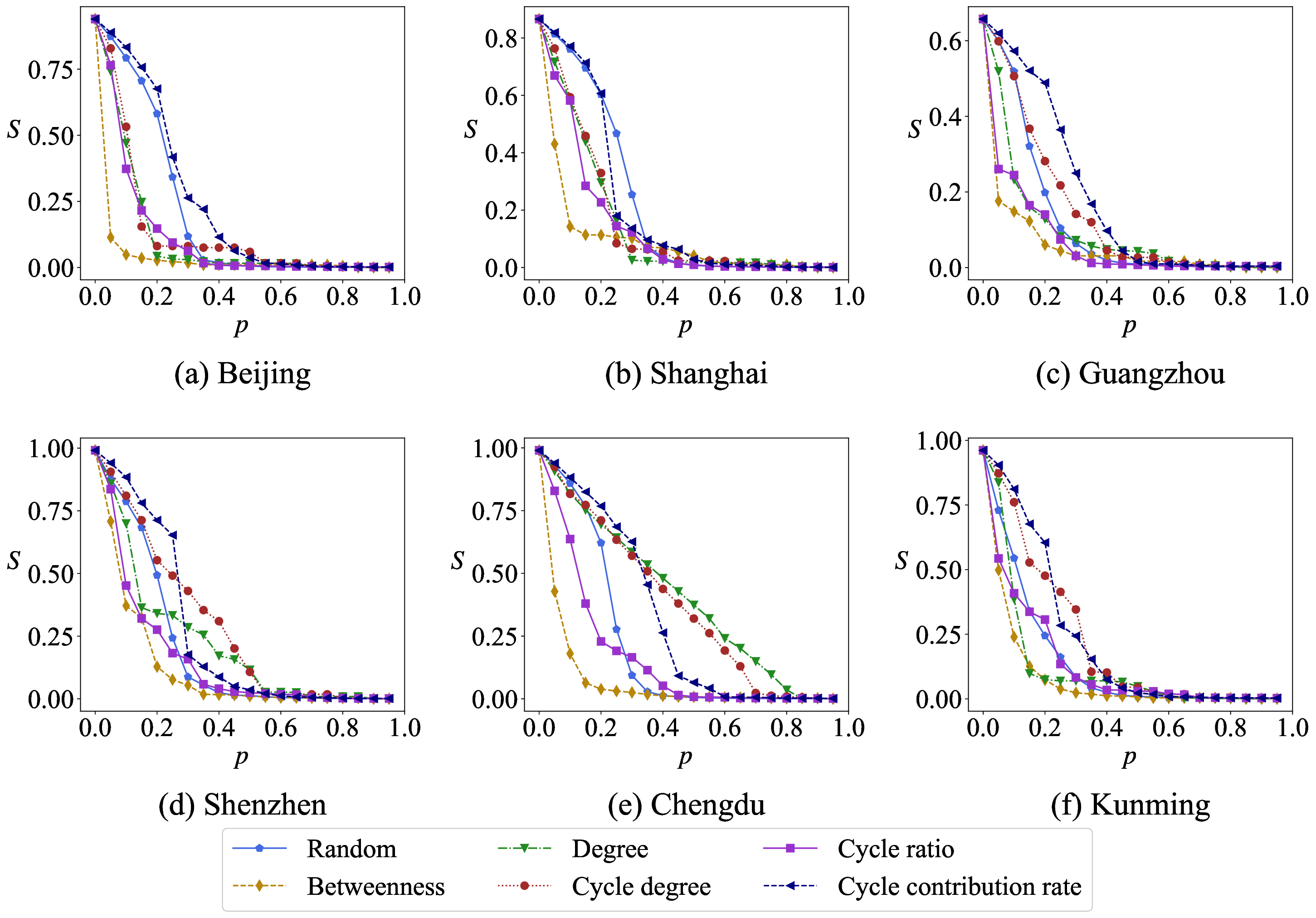}
\caption{The relative size of the maximum connected subgraph $S$ changes with the removing ratio $p$ under different removing strategies of nodes.}\label{fig10}
\end{figure}

In order to reduce the computing cost, we only select the second layer road network of six cities for robustness analysis. In urban road networks, the removal of nodes will affect the changes of network connectivity. Fig.~\ref{fig10} shows the relationship between the relative size of the maximum connected subgraph $S$ and the proportion of the removed nodes $p$ in the same network under six different removing strategies, removing nodes randomly and removing nodes according to degree, betweenness, cycle degree, cycle ratio and cycle contribution rate on network robustness. Obviously, the variation curves of the relative size of the maximum connected subgraph $S$ are different under different removing strategies of nodes. The betweenness centrality is relatively sensitive to disruptions in network connectivity, followed by the cycling rate, with the cycle contribution rate being the least sensitive. This may be attributed to the fact that nodes with higher cycle contribution rates are often located in the core areas of the road network, where they enjoy a high degree of connectivity, thus resulting in a weaker impact on network connectivity when disrupted.

\begin{figure}
	\centering
	\includegraphics[width=14cm]{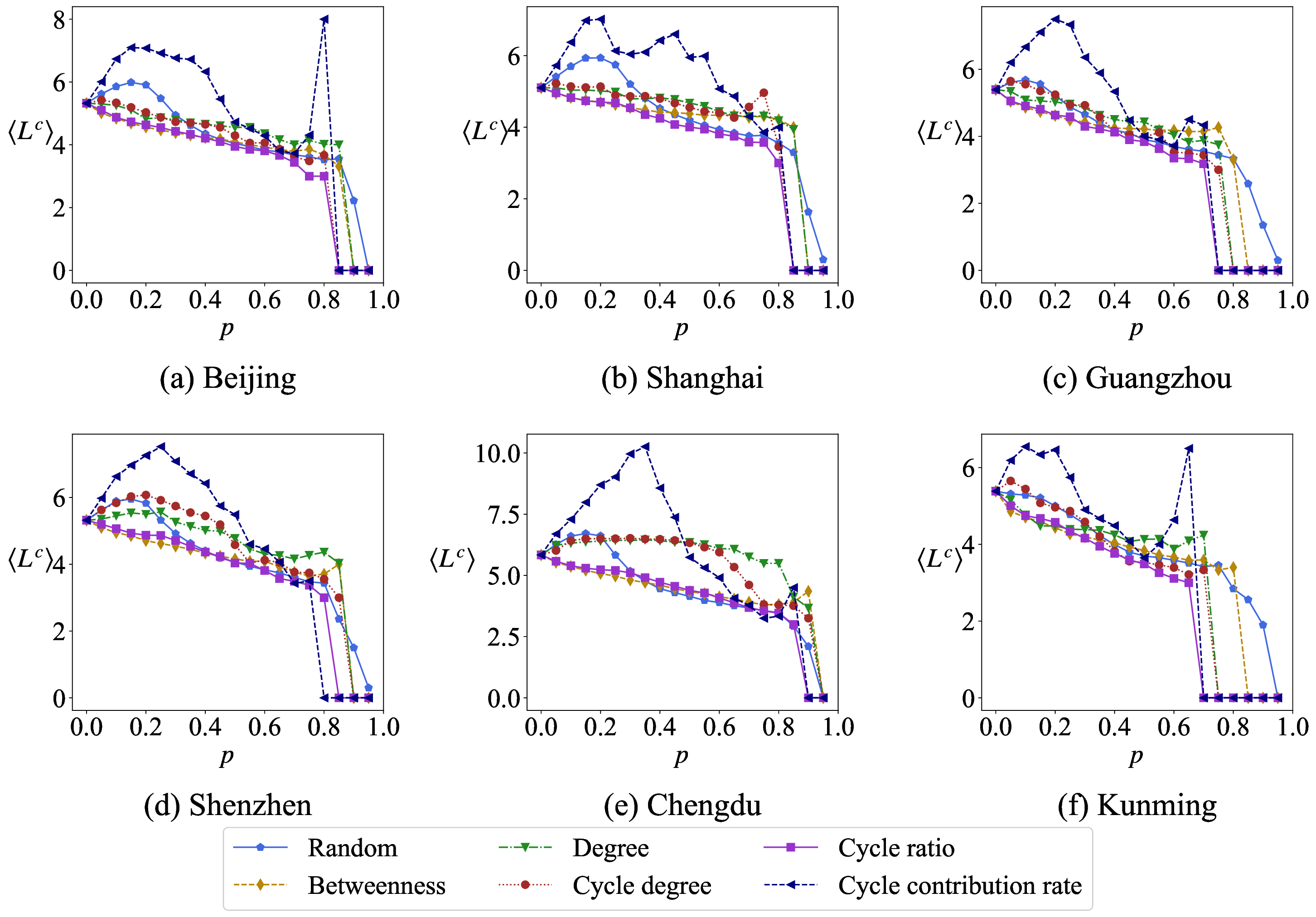}
\caption{The average length of the minimum cycle basis $\left \langle L^{c} \right \rangle$ changes with the removing ratio $p$ under different removing strategies.}\label{fig11}
\end{figure}

Removing nodes causes changes in the average length of the minimum cycle basis. Fig.~\ref{fig11} respectively shows the relationship between the average length of the minimum cycle basis $\left \langle L^{c} \right \rangle$ and the proportion of removed nodes $p$ in the same network under six different removing strategies, removing nodes randomly and removing nodes according to degree, betweenness, cycle degree, cycle ratio and cycle contribution rate on network robustness. Obviously, the variation curves of the average length of the minimum cycle basis $\left \langle L^{c} \right \rangle$ are different under different removing strategies of nodes. The change of $\left \langle L^{c} \right \rangle$ is more sensitive to cycle ratio than others. The cycle contribution rate has a relatively minor impact on $\left \langle L^{c} \right \rangle$, primarily because small cycles contribute more significantly to the cycle contribution rate, whereas larger cycles contribute more prominently to the cycling rate.

\begin{figure}
	\centering
	\includegraphics[width=14cm]{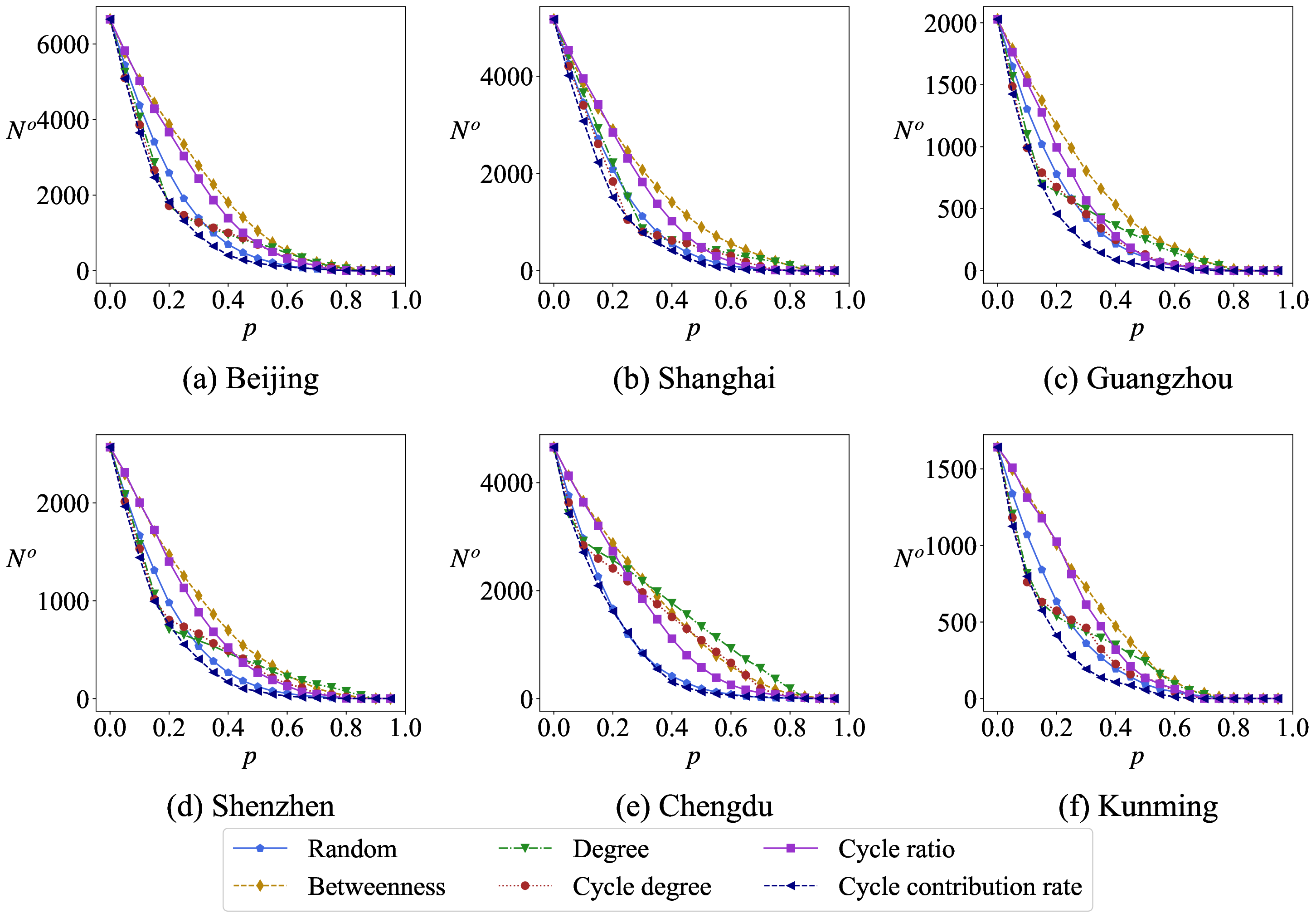}
\caption{The total number of cycles $N^{o}$ changes with the removing ratio $p$ under different removing strategies.}\label{fig12}
\end{figure}

The removal of network nodes will affect the change of the total number of the minimum cycle basis. Fig.~\ref{fig12} respectively shows the relationship between the total number of shortest cycles $N^{o}$ and the proportion of removed nodes $p$ in the same network under six different removing strategies, removing nodes randomly and removing nodes according to degree, betweenness, cycle degree, cycle ratio and cycle contribution rate on network robustness. Obviously, the variation curves of the total number of shortest cycles $N^{o}$ are different under different removing strategies of nodes. The change of $N^{o}$ is more sensitive to cycle contribution rate than others. In a road network of a given area, an increase in the number of cycles indicates superior network connectivity, providing more routing options for navigation and enhancing convenience in daily life. That is to say, the cycle contribution rate characterizes the level of connectivity density within a specific region, functioning as a valuable indicator for assessing the polycentric character of road networks.

In summary, when studying the robustness of the network, with the increase of the number of removed nodes, the betweenness affects the relative size change of the maximum connected subgraph significantly. The cycle ratio affects the average length of the minimum cycle basis significantly. The cycle contribution rate affects the total number of cycles significantly.

\subsection{Analyzing cycle-based dual network}
\begin{figure}[htbp]
\begin{center}
\subfigure[]{
		\includegraphics[width=3cm]{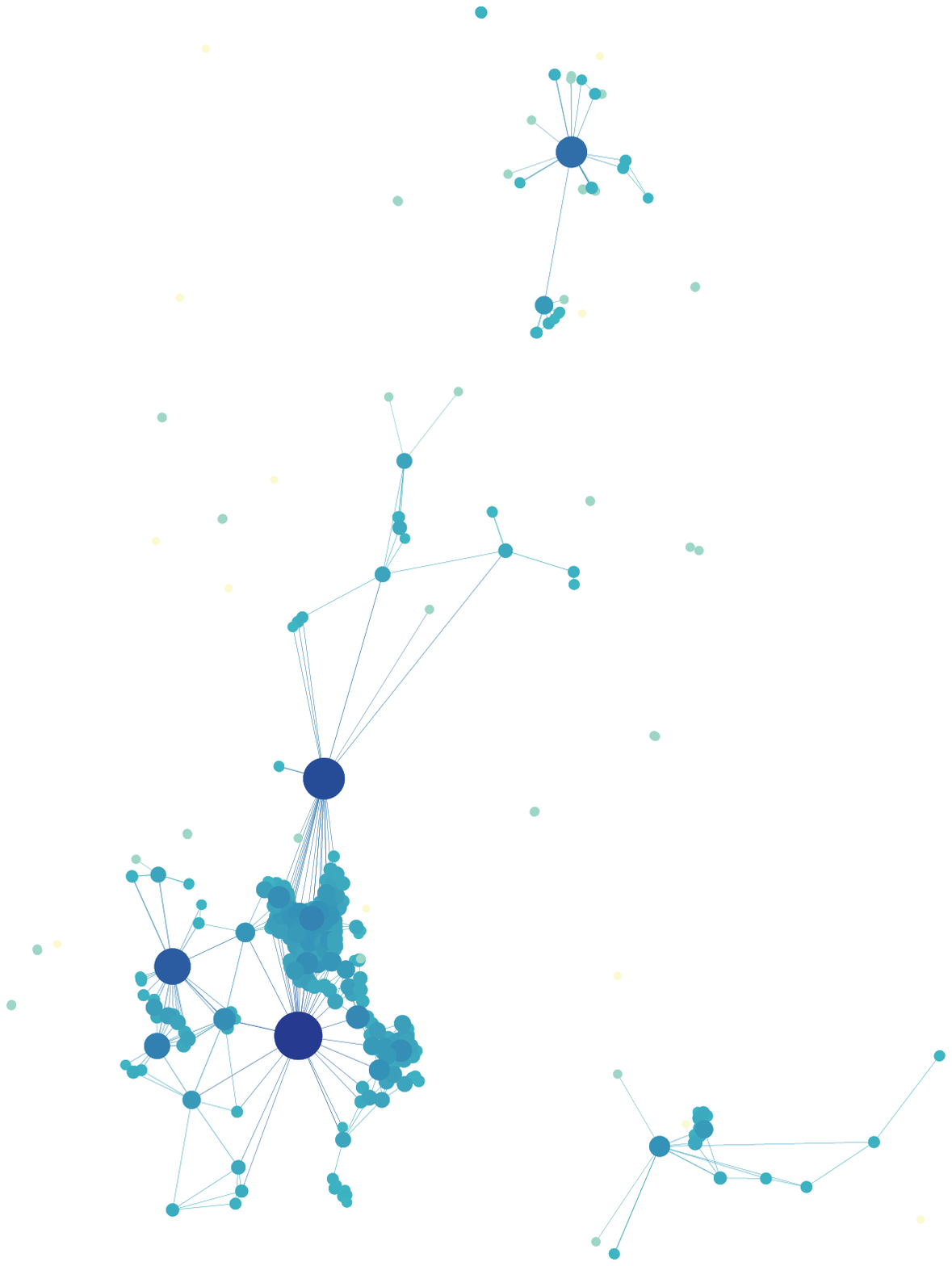}
		\label{fig13a}  }
\subfigure[]{
		\includegraphics[width=3cm]{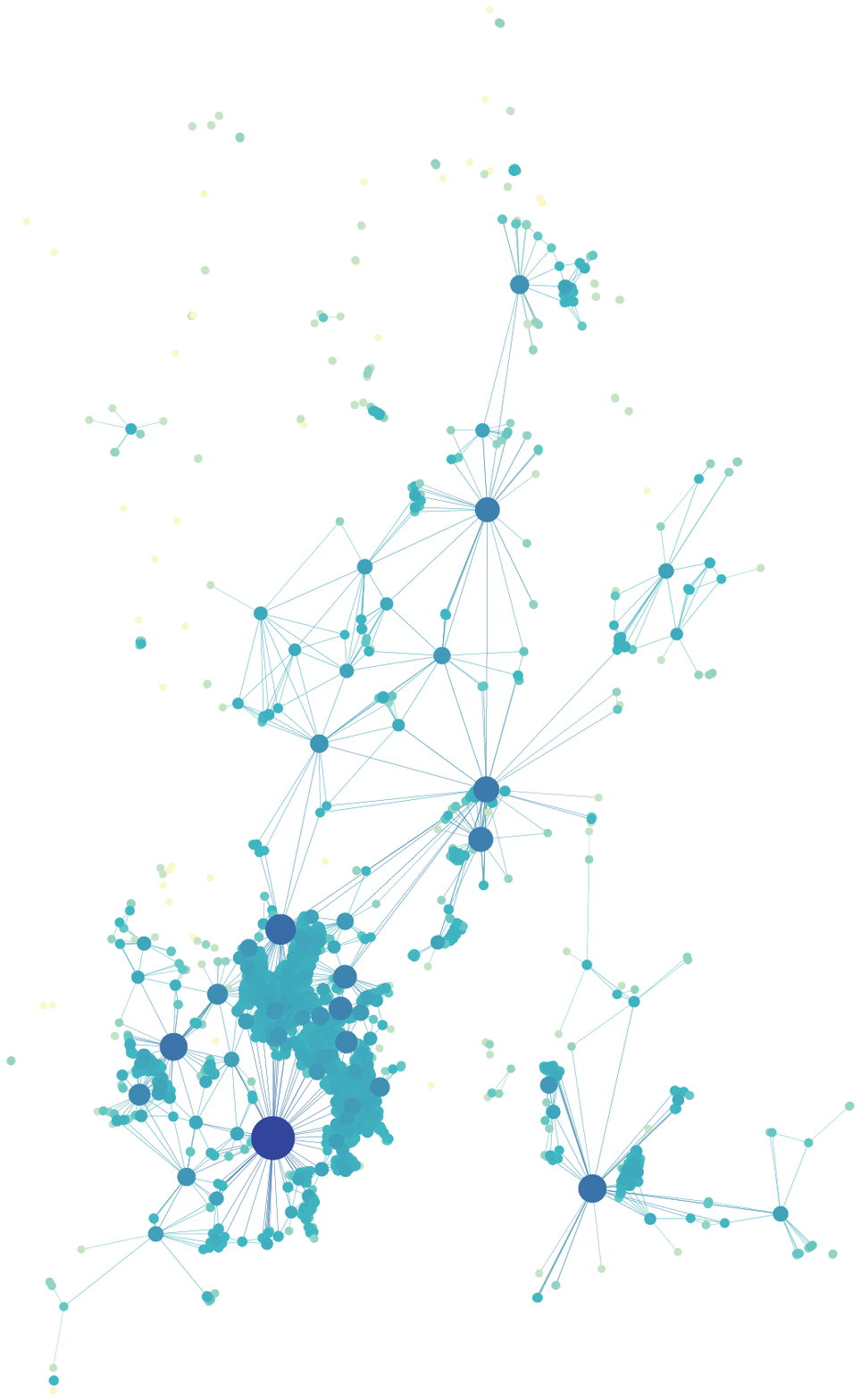}
		\label{fig13b} }
\subfigure[]{
		\includegraphics[width=3cm]{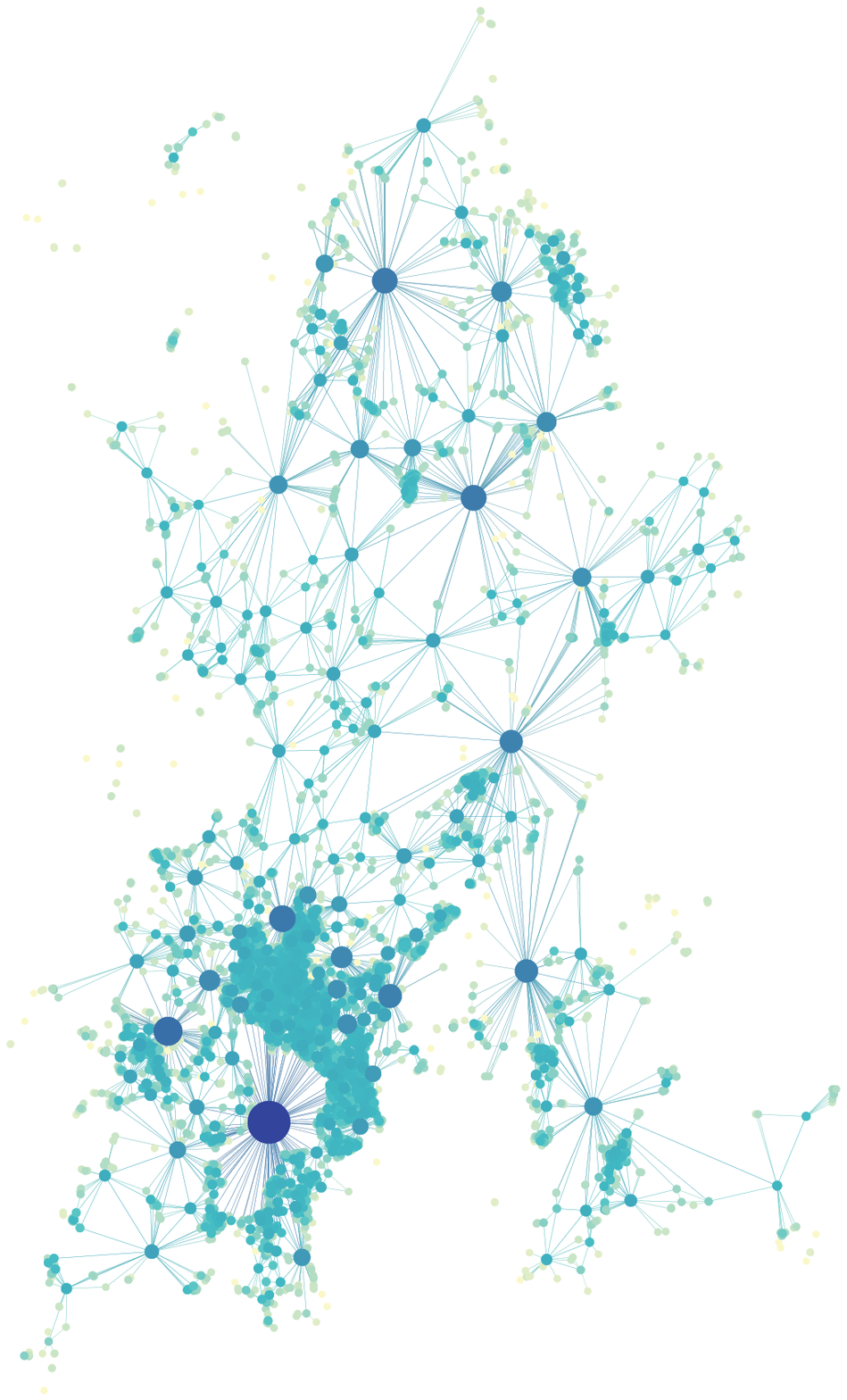}
		\label{fig13c} }
\subfigure[]{
		\includegraphics[width=3cm]{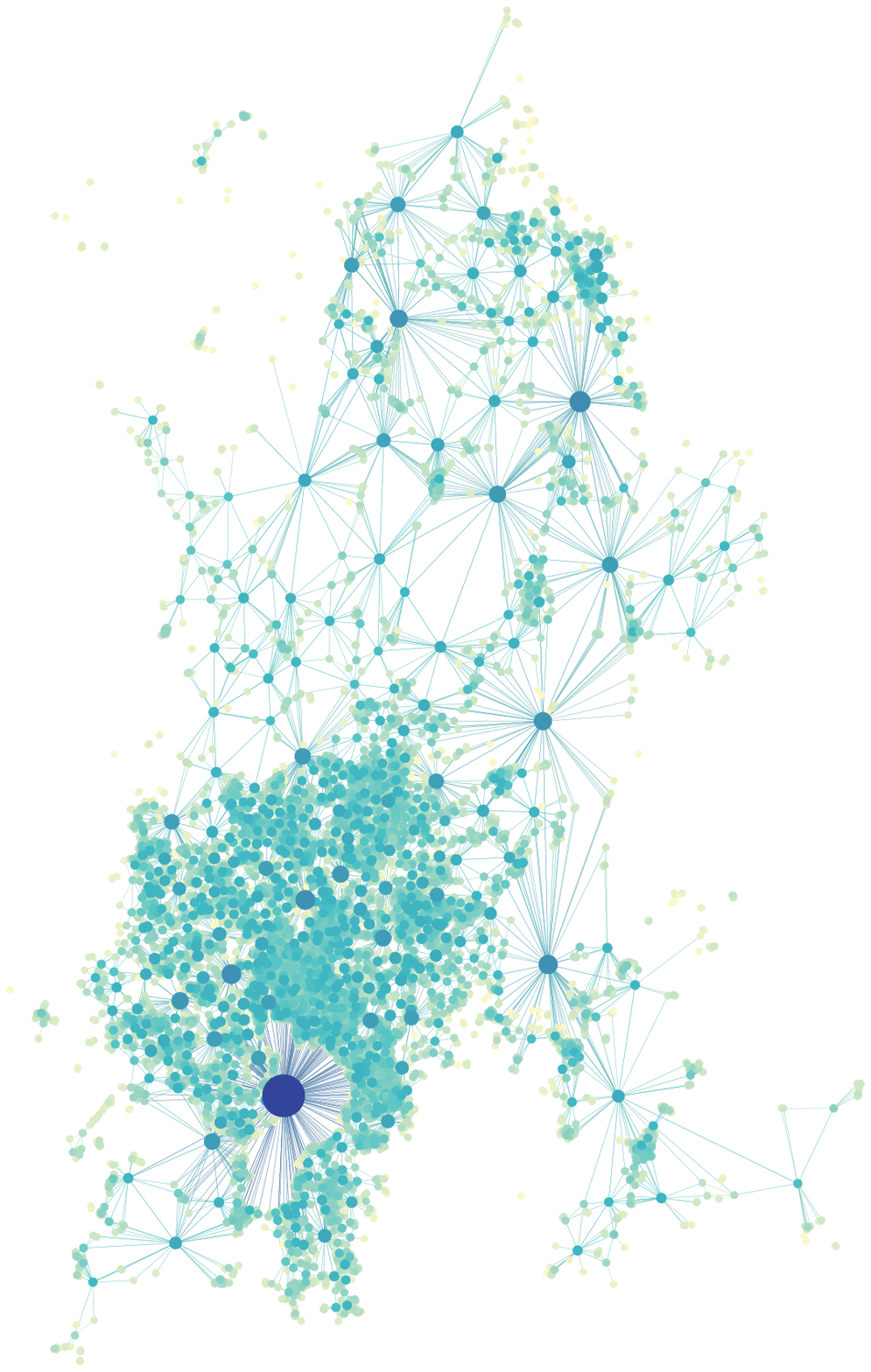}
		\label{fig13d}  }
\subfigure[]{
		\includegraphics[width=3cm]{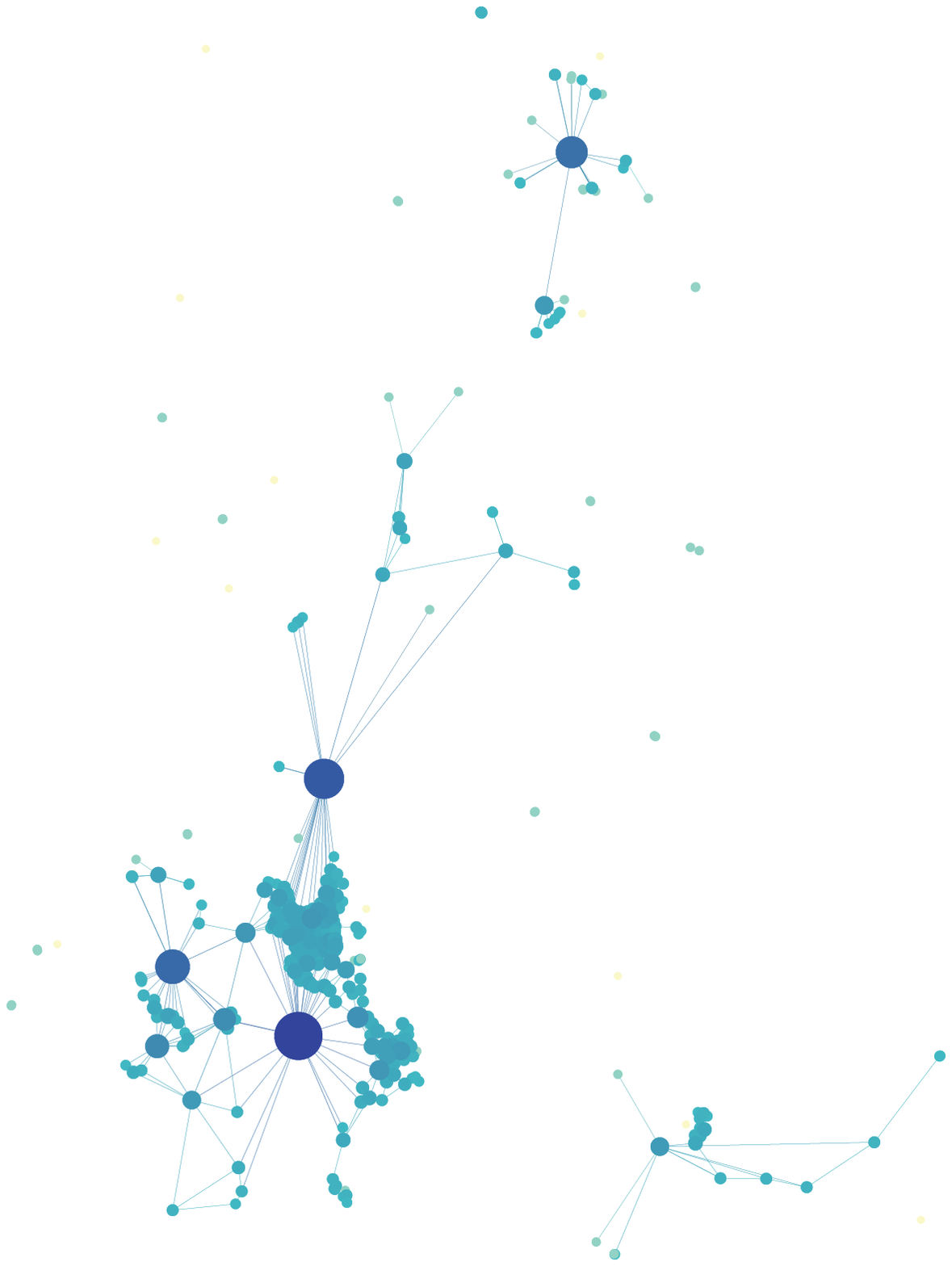}
		\label{fig13e}  }
\subfigure[]{
		\includegraphics[width=3cm]{figure/dual2_1.eps}
		\label{fig13f} }
\subfigure[]{
		\includegraphics[width=3cm]{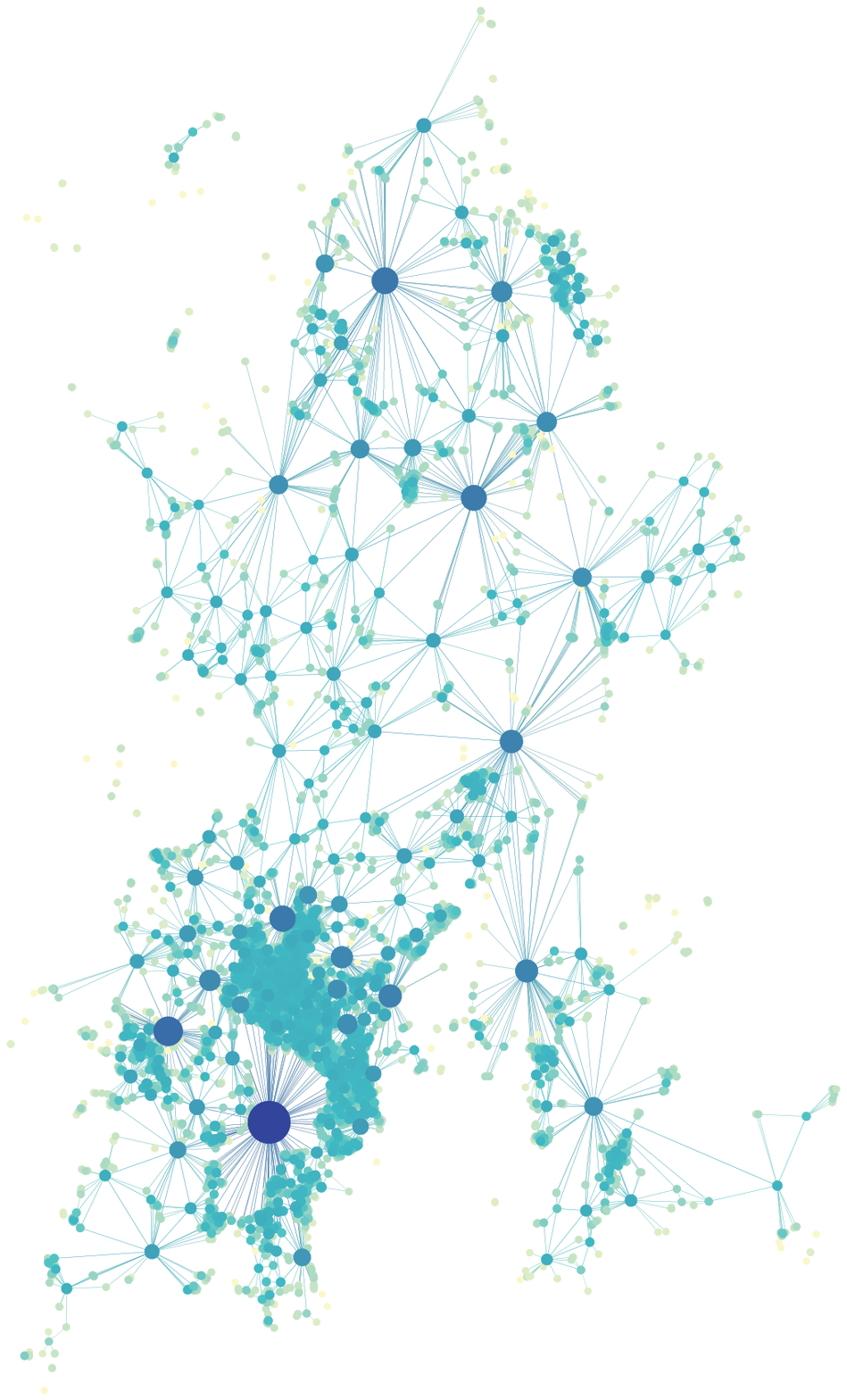}
		\label{fig13g} }
\subfigure[]{
		\includegraphics[width=3cm]{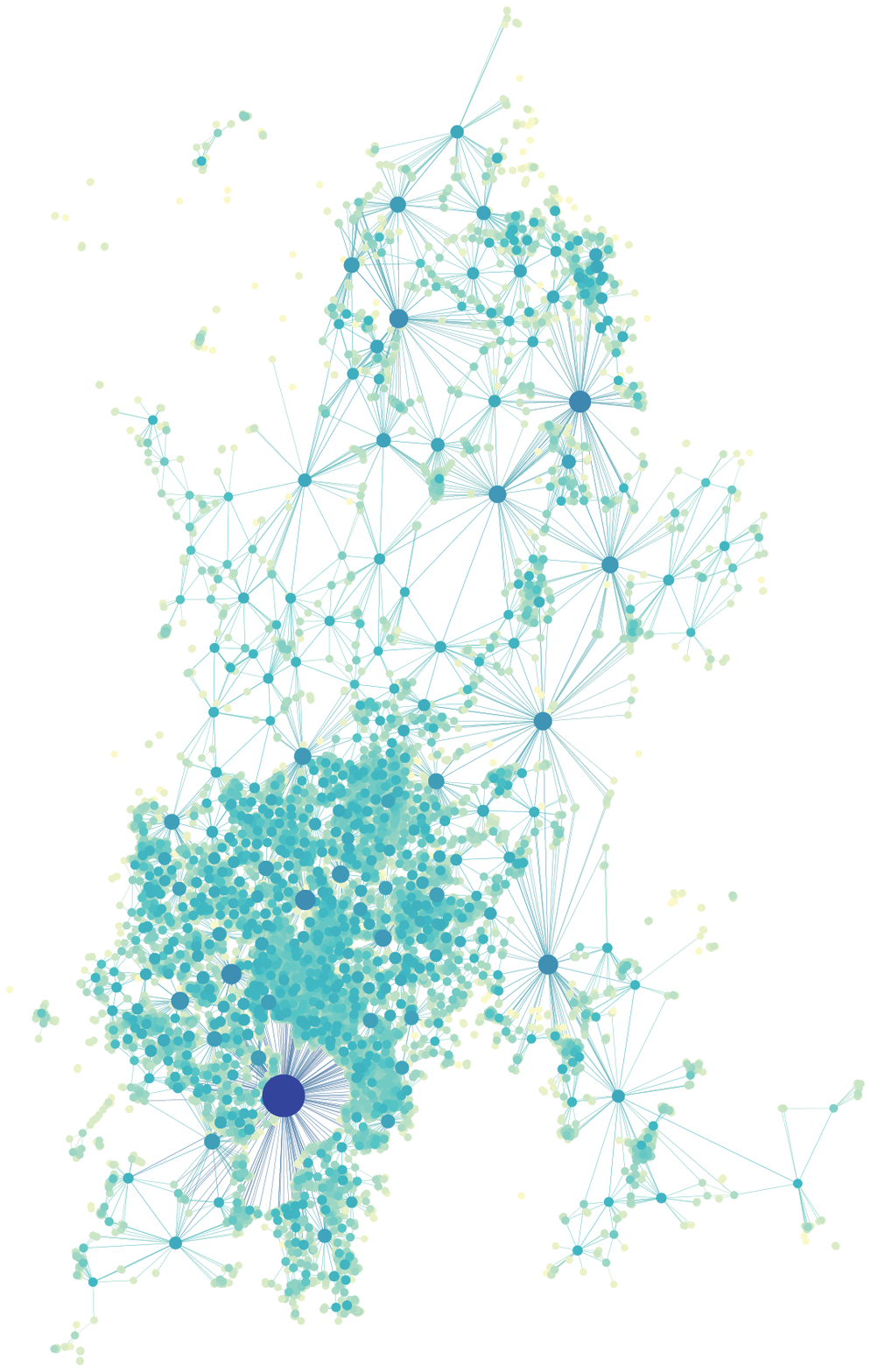}
		\label{fig13h}  }
\caption{Different layers of cycle-based dual network of Kunming. (a)-(d) are the first-fourth layer in sequence and share one node, (e)-(h) are the first-fourth layer in sequence and share one edge.}
\label{fig13}
\end{center}
\end{figure}

Cycle-based dual network is constructed by treating the smallest cycles as nodes and establishing connections based on shared nodes between cycles. Based on it, we establish two types of cycle-based dual networks, that is, the cycle-based dual network in which the cycles share at least one node and the cycle-based dual network in which the cycles share at least one edge. Taking four layers of Kunming road network as an example, Fig.~\ref{fig13} shows the schematic diagrams of two types of cycle-based dual network. As depicted in the graph, the degree distribution displays power-law properties, characterized by a small proportion of nodes possessing significantly higher degree values. Notably, the node exhibiting the maximum degree value corresponds to the circular roadway encompassing Dianchi Lake.

\begin{figure}
	\centering
	\includegraphics[width=14cm]{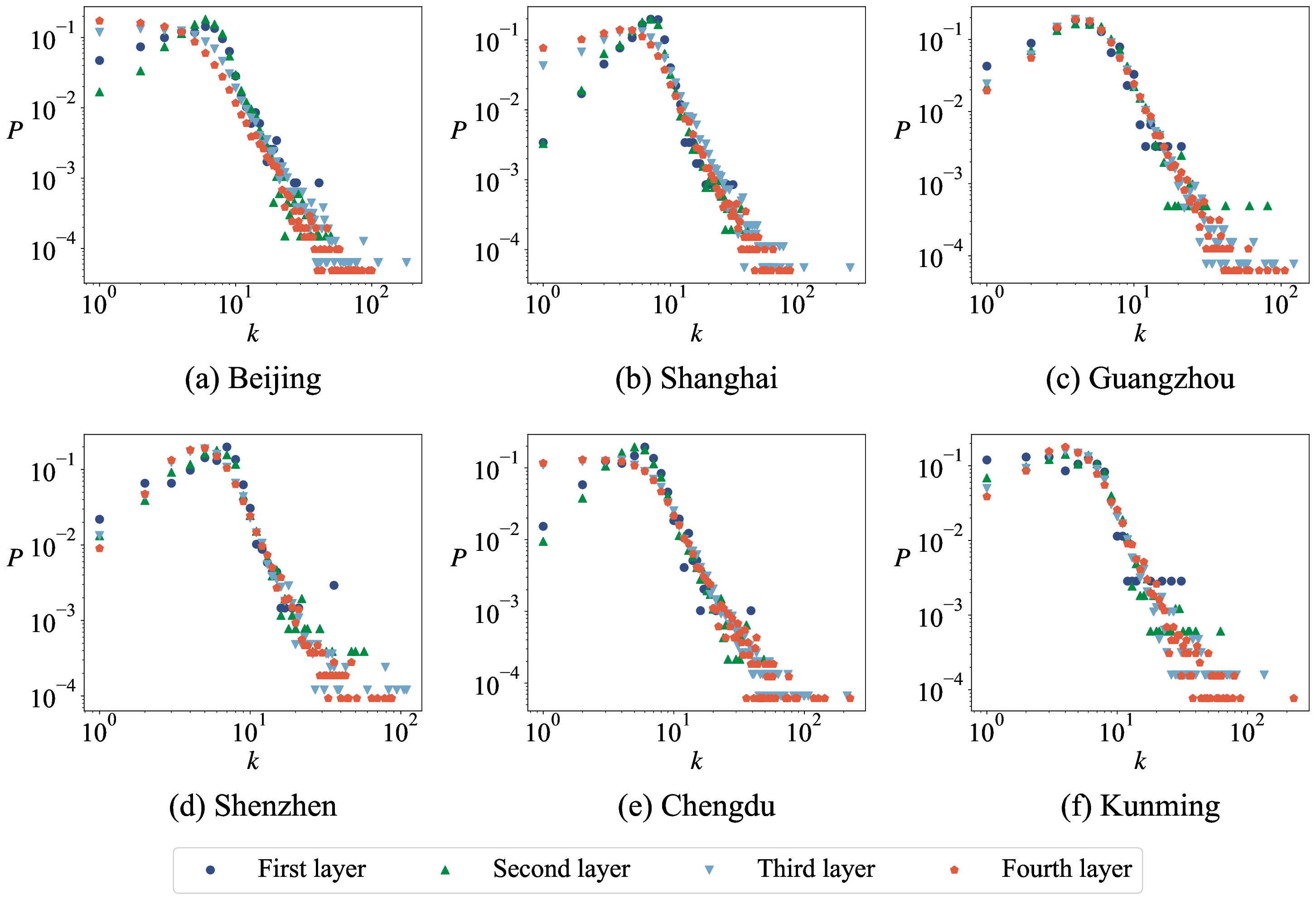}
\caption{The degree distribution of the cycle-based dual network sharing one node.}\label{fig14}
\end{figure}

\begin{figure}
	\centering
	\includegraphics[width=14cm]{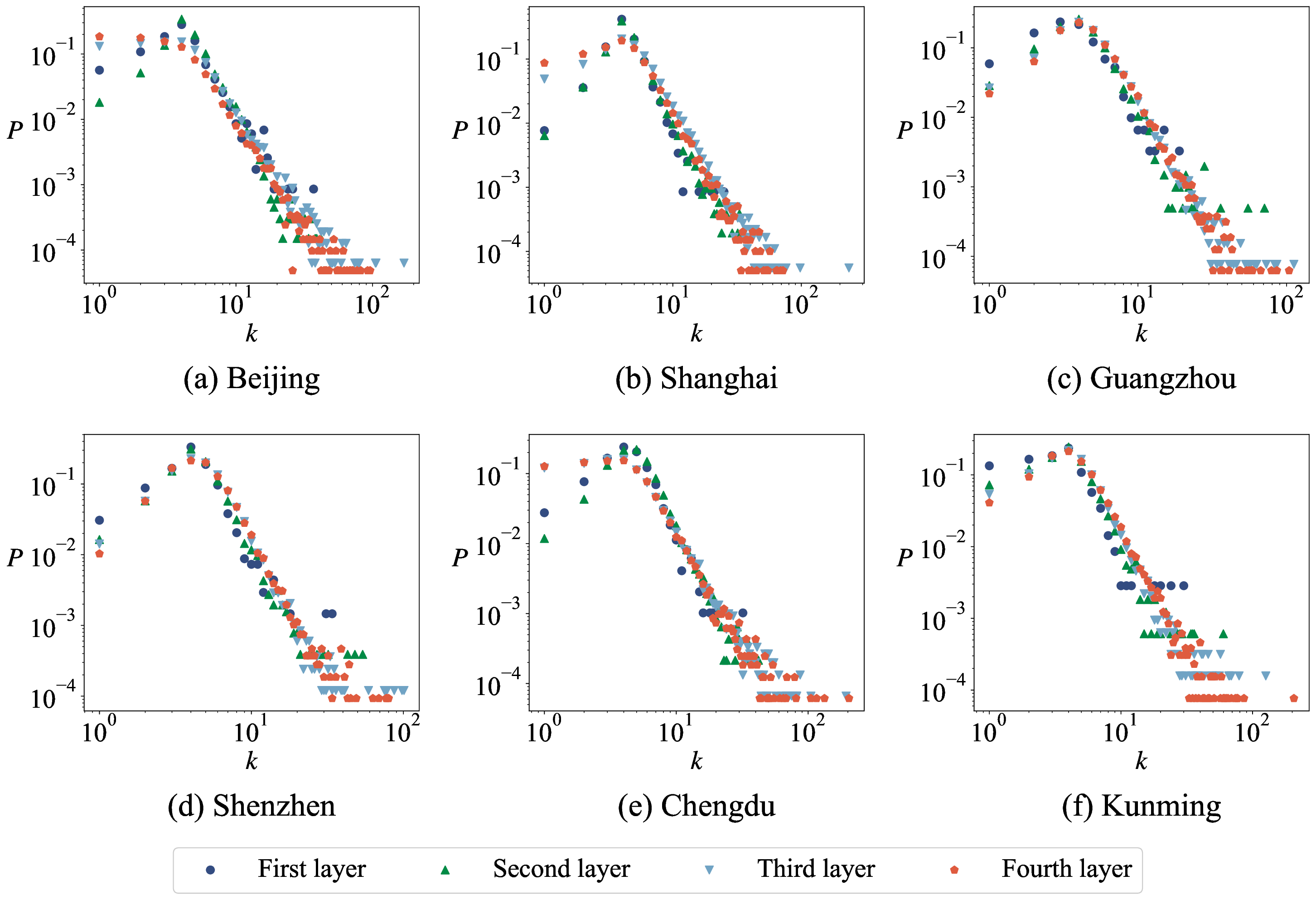}
\caption{The degree distribution of the cycle-based dual network sharing one edge.}\label{fig15}
\end{figure}

The degree distribution of two kinds of cycle-based dual networks of six cities is showed in Fig.~\ref{fig14} and Fig.~\ref{fig15}, which follows power-law distribution. In cycle-based dual network in which the cycles share at least one node, the degree distribution demonstrates power-law behavior when the degree value surpasses 6. Conversely, when the degree value is 6 or below, variations arise across different layers and cities. Similarly, the cycle-based dual network in which the cycles share at least one edge exhibit analogous characteristics, albeit with a modified threshold degree value of 4.

Table~\ref{Tab6} and Table~\ref{Tab7} illustrate the topological characteristics including the number of nodes $N$, the number of edges $M$, average degree $\langle k \rangle$, average path length $\langle d \rangle$, clustering coefficient $\langle C \rangle$ and network diameter $D$ of two kinds of cycle-based dual networks of six cities, corresponding to Fig.~\ref{fig14} and Fig.~\ref{fig15} respectively. The results show that two kinds of cycle-based dual networks possess higher average degree and clustering coefficient, and a shorter average path length and network diameter, showing more obvious small-world property compared with the results of urban road networks as shown in Table~\ref{Tab2}.

\begin{table}[htbp]
\centering
\caption{The topological characteristics of cycle-based dual network sharing one node. $N^{v}$: the number of nodes; $M^{v}$: the number of edges; $\langle k^{v} \rangle$: average degree; $\mathcal{C}^{v}$: the number of connected subgraphs; $N_{max}^{v}$: the number of nodes in the maximal connected subgraph; $\langle d^{v} \rangle$: average path length of the maximal connected subgraph; $\langle C^{v} \rangle$: clustering coefficient of the maximal connected subgraph; $D^{v}$: network diameter of the maximal connected subgraph.}
\label{Tab6}
\footnotesize
\begin{tabular}{@{}llcccccccc}
\br
\textbf{City} & \textbf{Level} & $\bm{N^{v}}$ & $\bm{M^{v}}$ & $\bm{\langle k^{v} \rangle}$ & $\bm{\mathcal{C}^{v}}$ & $\bm{N_{max}^{v}}$ & $\bm{\langle d^{v} \rangle}$ & $\bm{\langle C^{v} \rangle}$ & $\bm{D^{v}}$ \\
\mr
\multirow{4}{*}{Beijing} & First layer & 1167 & 3428 & 5.87 & 43 & 994 & 9.88 & 0.54 & 28 \\
 & Second layer & 6656 & 21285 & 6.40 & 91 & 6435 &  16.27 & 0.52 & 46 \\
 & Third layer & 27170 & 87385 & 6.43 & 175 & 24804 & 27.21 & 0.55 & 79 \\
 & Fourth layer & 42887 & 133913 & 6.24 & 269 & 39864 & 33.71 & 0.55 & 89 \\
 \mr
\multirow{4}{*}{Shanghai} & First layer & 1175 & 4082 & 6.95 & 3 & 1102 & 9.97 & 0.51 & 23 \\
 & Second layer & 5173 & 17437 & 6.74 & 12 & 4605 & 15.56 & 0.51 & 39 \\
 & Third layer & 23022 & 79227 & 6.88 & 42 & 19634 & 22.40 & 0.54 & 59 \\
 & Fourth layer & 29936 & 99225 & 6.63 & 50 & 26080 & 27.35 & 0.54 & 82 \\
  \mr
\multirow{4}{*}{Guangzhou} & First layer & 305 & 764 & 5.01 & 13 & 159 & 6.19 & 0.57 & 17 \\
 & Second layer & 2029 & 5758 & 5.68 & 42 & 1216 & 7.01 & 0.57 & 18 \\
 & Third layer & 13019 & 36649 & 5.63 & 193 & 8149 & 12.69 & 0.56 & 43 \\
 & Fourth layer & 16029 & 45634 & 5.69 & 188 & 10531 & 16.91 & 0.56 & 52 \\
  \mr
\multirow{4}{*}{Shenzhen} & First layer & 684 & 2132 & 6.23 & 9 & 594 & 6.82 & 0.53 & 16 \\
 & Second layer & 2569 & 7979 & 6.21 & 14 & 2122 & 10.69 & 0.53 & 26 \\
 & Third layer & 8312 & 24304 & 5.85 & 36 & 8224 & 16.00 & 0.54 & 46 \\
 & Fourth layer & 10753 & 31256 & 5.81 & 39 & 10617 & 19.17 & 0.55 & 61 \\
  \mr
\multirow{4}{*}{Chengdu} & First layer & 979 & 2869 & 5.86 & 11 & 915 & 9.06 & 0.52 & 29 \\
 & Second layer & 4658 & 13834 & 5.94 & 26 & 4594 & 13.08 & 0.53 & 32 \\
 & Third layer & 26618 & 87331 & 6.56 & 144 & 26288 & 27.48 & 0.55 & 77 \\
 & Fourth layer & 28755 & 93768 & 6.52 & 153 & 28396 & 28.76 & 0.55 & 73 \\
  \mr
\multirow{4}{*}{Kunming} & First layer & 351 & 830 & 4.73 & 30 & 256 & 5.66 & 0.51 & 13 \\
 & Second layer & 1642 & 4432 & 5.40 & 80 & 1343 & 7.69 & 0.53 & 22 \\
 & Third layer & 6340 & 17482 & 5.51 & 173 & 5951 & 10.91 & 0.55 & 29 \\
 & Fourth layer & 13051 & 36717 & 5.63 & 266 & 12407 & 13.23 & 0.57 & 37 \\
\br
\end{tabular}
\end{table}

\begin{table}[htbp]
\centering
\caption{The topological characteristics of cycle-based dual network sharing one edge. $N^{e}$: the number of nodes; $M^{e}$: the number of edges; $\langle k^{e} \rangle$: average degree; $\mathcal{C}^{e}$: the number of connected subgraphs; $N_{max}^{e}$: the number of nodes in the maximal connected subgraph; $\langle d^{e} \rangle$: average path length of the maximal connected subgraph; $\langle C^{e} \rangle$: clustering coefficient of the maximal connected subgraph; $D^{e}$: network diameter of the maximal connected subgraph.}
\label{Tab7}
\footnotesize
\begin{tabular}{@{}llcccccccc}
\br
\textbf{City} & \textbf{Level} & $\bm{N^{e}}$ & $\bm{M^{e}}$ & $\bm{\langle k^{e} \rangle}$ & $\bm{\mathcal{C}^{e}}$ & $\bm{N_{max}^{e}}$ & $\bm{\langle d^{e} \rangle}$ & $\bm{\langle C^{e} \rangle}$ & $\bm{D^{e}}$ \\
\mr
\multirow{4}{*}{Beijing} & First layer & 1167 & 2560 & 4.39 & 45 & 989 & 10.80 & 0.39 & 31 \\
 & Second layer & 6656 & 16401 & 4.93 & 94 & 6432 & 17.74 & 0.40 & 52 \\
 & Third layer & 27170 & 77248 & 5.69 & 185 & 24780 & 28.37 & 0.49 & 83 \\
 & Fourth layer & 42887 & 122138 & 5.70 & 287 & 39832 & 35.25 & 0.51 & 93 \\
 \mr
\multirow{4}{*}{Shanghai} & First layer & 1175 & 2670 & 4.54 & 4 & 1102 & 12.10 & 0.28 & 28 \\
 & Second layer & 5173 & 12509 & 4.84 & 14 & 4597 & 17.57 & 0.35 & 47 \\
 & Third layer & 23022 & 66741 & 5.80 & 44 & 19624 & 24.06 & 0.46 & 67 \\
 & Fourth layer & 29936 & 86875 & 5.80 & 53 & 26078 & 28.76 & 0.48 & 83 \\
 \mr
\multirow{4}{*}{Guangzhou} & First layer & 305 & 606 & 3.97 & 15 & 132 & 5.03 & 0.41 & 13 \\
 & Second layer & 2029 & 4772 & 5.68 & 44 & 1214 & 7.40 & 0.46 & 19 \\
 & Third layer & 13019 & 33701 & 5.18 & 198 & 8141 & 13.10 & 0.52 & 45 \\
 & Fourth layer & 16029 & 42196 & 5.27 & 197 & 10515 & 17.72 & 0.52 & 56 \\
 \mr
\multirow{4}{*}{Shenzhen} & First layer & 684 & 1519 & 4.44 & 9 & 594 & 7.93 & 0.34 & 21 \\
 & Second layer & 2569 & 6302 & 4.91 & 15 & 2122 & 11.73 & 0.41 & 28 \\
 & Third layer & 8312 & 22207 & 5.34 & 41 & 8217 & 16.41 & 0.49 & 46 \\
 & Fourth layer & 10753 & 28996 & 5.39 & 41 & 10616 & 19.47 & 0.51 & 61 \\
 \mr
\multirow{4}{*}{Chengdu} & First layer & 979 & 2348 & 4.80 & 12 & 912 & 10.13 & 0.43 & 33 \\
 & Second layer & 4658 & 12709 & 5.46 & 30 & 4589 & 13.54 & 0.49 & 33 \\
 & Third layer & 26618 & 76593 & 5.76 & 156 & 26268 & 29.42 & 0.49 & 81 \\
 & Fourth layer & 28755 & 83576 & 5.81 & 161 & 28385 & 30.39 & 0.50 & 76 \\
 \mr
\multirow{4}{*}{Kunming} & First layer & 351 & 638 & 3.64 & 30 & 256 & 6.23 & 0.38 & 15 \\
 & Second layer & 1642 & 3612 & 4.40 & 83 & 1309 & 8.05 & 0.43 & 22 \\
 & Third layer & 6340 & 15608 & 4.92 & 180 & 5943 & 11.34 & 0.50 & 31 \\
 & Fourth layer & 13051 & 34001 & 5.21 & 272 & 12400 & 13.47 & 0.54 & 38 \\
 \br
\end{tabular}
\end{table}

In summary, when analyzing urban road networks with the minimum cycle basis serving as the fundamental component, the network displays both small-world and scale-free properties. Notably, the clustering coefficient hovering around 0.5 indicates a specific structure, while the degree distribution of the network conforms to a power-law pattern. As widely recognized, the small-world property signifies a higher degree of interconnectedness and intimacy among nodes, whereas the scale-free property emphasizes the heterogeneity in degree values among nodes, which facilitates the rapid dissemination of information and the distribution of resources. These findings offer a fresh perspective on road network research and implicitly illustrate how the prevalent ring highways in cities, functioning as significant nodes in the cycle-based dual network, can significantly enhance urban living convenience.

\section{Conclusions}\label{S4}
This study focuses on six cities in China(Beijing, Shanghai, Guangzhou, Shenzhen, Chengdu and Kunming) to explore the topological characteristics, the centrality of node and robustness of urban road networks based on motif and the minimum cycle basis. To display the hierarchical structure of urban roads and reduce the computational complexity, we select four layers of urban road networks, including the first layer(main roads), the second layer (main roads and secondary trunk roads), the third layer (main roads, secondary trunk roads and branch roads) and the fourth layer (main roads, secondary trunk roads, branch roads and internal roads).

Firstly, this study initially conducts motif analysis on the urban road networks of Beijing, Shanghai, Guangzhou, Shenzhen, Chengdu, and Kunming, revealing that motifs containing cycles exhibit positive Z-scores, indicating their prevalence exceeds that of equivalent-degree sequences in random networks; Leveraging this finding, we undertake topological and robustness analyses of urban road networks using the minimum cycle basis as a fundamental element. Intriguingly, we observe that the length distribution of the minimum cycle basis in urban road networks follows a power-law distribution; To investigate node importance within urban road networks, we introduce the concept of cycle contribution rate(the importance of a node depends not only on the number of cycles it is associated with but also on the length of these cycles) and compare the results with node degree, cycle degree(the number of cycles associated with a node), and cycle ratio. The findings indicate that nodes with higher cycle degrees typically exhibit higher cycle ratios, while those associated with a greater number of smaller cycles tend to demonstrate higher cycle contributions, underscoring the advantageous role of cycle contribution in identifying central areas within road networks; Subsequently, we adopt the relative size of the maximum connected subgraph, the number of cycles in the network, and the average length of cycles as observations to verify the validity of cycle contribution rate to node importance detection. The robustness under removing nodes randomly and removing nodes according to degree, betweenness, cycle degree, cycle ratio are used to compare with that under the cycle contribution rate. The results reveal that with an increasing number of removed nodes, betweenness centrality significantly impacts the relative size of the maximum connected subgraph, while cycle ratio notably affects the average length of cycles, and cycle contribution rate substantially influences the number of cycles; Finally, we construct two types of cycle-based dual networks for urban road networks by representing cycles as nodes and establishing edges between two cycles sharing a common node and edge respectively. The dual network exhibits small-world and scale-free properties, manifested through a power-law degree distribution, high clustering coefficients, and short average path lengths.

The adoption of the minimum cycle basis as a fundamental unit for analyzing urban road networks offers a fresh perspective for associated research endeavors, presenting the following promising applications: (1) In the process of constructing road network generation models, one may incorporate the power-law distribution characteristics of the lengths and areas associated with the minimum cycle basis; (2) By comparing the topological attributes of the minimum cycle basis across diverse cities, existing road networks can be enhanced and optimized; (3) Potential applications can be explored, such as utilizing the cycle contribution rate, as introduced in this paper, to delve into the multi-centrality of networks. This metric demonstrates notable performance in both node differentiation and spatial distribution. Future endeavors may further explore related application scenarios, including community detection and link prediction grounded in the minimum cycle basis. Moreover, endeavoring to establish a comprehensive road network description and analysis framework based on the minimum cycle basis will undoubtedly aid in addressing various road-related challenges.

\section*{Acknowledgments}
This work was supported by Yunnan Fundamental Research Projects (grant NO. 202401AT070359).

\section*{Data availability statement}
The data that support the findings of this study are available upon reasonable request from the corresponding author.

\section*{Computer code availability statement}
The codes and partial data have been posted at https://github.com/kust-yangbo/Urban-Road-Networks/tree/master.

\end{document}